\documentclass[printer]{aa}
\usepackage[varg]{txfonts}
\usepackage{natbib, amsmath, amssymb, amsfonts, graphicx,xcolor}
\usepackage{mathtools}
\usepackage[citecolor=blue, linkcolor=blue, urlcolor = black, colorlinks = true]{hyperref}

\title{The effect of the centrifugal acceleration on period spacings \\of gravito-inertial modes in intermediate-mass stars}

\author{J. Henneco \inst{1}
\and T. Van Reeth \inst{1}
\and V. Prat \inst{2}
\and S. Mathis \inst{2}
\and J.\,S.\,G. Mombarg \inst{1}
\and C. Aerts \inst{1,3,4}}

\institute{Institute of Astronomy, KU Leuven, Celestijnenlaan 200D, 3001 Leuven, Belgium\\ \email{jan.henneco@protonmail.com}
\and AIM, CEA, CNRS, Université Paris-Saclay, Université Paris Diderot, Sorbonne Paris Cité, 91191 Gif-sur-Yvette, France
\and Department of Astrophysics, IMAPP, Radboud University Nijmegen, PO Box 9010, 6500 GL Nijmegen, the Netherlands
\and Max Planck Institute for Astronomy, Koenigstuhl 17, 69117 Heidelberg, Germany}

\titlerunning{Effect of centrifugal deformation on g-mode pulsations}
\authorrunning{Henneco et al.}

\abstract{The \textit{Kepler} and TESS space telescopes delivered high-precision, long-duration photometric time series for hundreds of main-sequence stars, revealing  their numerous gravito-inertial (\emph{g}) pulsation modes. This high precision allows us to evaluate increasingly detailed theoretical stellar models. Recent theoretical work extended the traditional approximation of rotation, a framework to evaluate the effect of the Coriolis acceleration on g-modes, to include the effects of the centrifugal acceleration in the approximation of slightly deformed stars, which so far had mostly been neglected in asteroseismology. This extension of the traditional approximation was conceived by rederiving the traditional approximation in a centrifugally deformed, spheroidal coordinate system.}{We explore the effect of the centrifugal acceleration on g~modes and assess its detectability in space-based photometric observations.}{We implement the new theoretical framework to calculate the centrifugal deformation of precomputed 1D spherical stellar structure models and compute the corresponding g-mode frequencies, assuming uniform rotation. The framework is evaluated for a grid of stellar structure models covering a relevant parameter space for observed g-mode pulsators.}{The centrifugal acceleration modifies the effect of the Coriolis acceleration on g~modes,  narrowing the equatorial band in which they are trapped. Furthermore, the centrifugal acceleration causes the pulsation periods and period spacings of the most common g~modes (prograde dipole modes and r~modes) to increase with values similar to the observational uncertainties of the measured period spacing values in \textit{Kepler} and TESS data.}{The effect of the centrifugal acceleration on g~modes is formally detectable in modern space photometry. Implementation of the used theoretical framework in stellar structure and pulsation codes will allow for more precise asteroseismic modelling of centrifugally deformed stars, in order to assess its effect on mode excitation, -trapping and -damping.}

\begin{document}
\maketitle

%%%%%%%%%%%%%%%%%%%%%%%%%%%%%%%%% INTRODUCTION %%%%%%%%%%%%%%%%%%%%%%%%%%%%%%%%%
\section{Introduction}
Over the last decade, there have been major advancements in observational asteroseismology. Thanks to space missions such as CoRoT \citep[Convection, Rotation and planetary Transits;][]{auvergne2009}, \textit{Kepler} \citep{borucki2009}, the BRITE-Constellation \citep[BRIght Target Explorer Constellation;][]{weiss2014} and TESS \citep[Transiting Exoplanet Survey Satellite;][]{ricker2014}, long-time-base, high-cadence, high-precision photometric light curves are now available for hundreds of thousands of stars. This has resulted in the detection and identification of g-mode oscillations, which have buoyancy as the main restoring force, in hundreds of $\gamma$\,Doradus ($\gamma$\,Dor) and slowly-pulsating B-type (SPB) stars \citep[e.g.,][Pedersen et al., submitted]{tkachenko2013, vanreeth2015, papics2017, christophe2018, li2019a, li2019b, li2020}. 

$\gamma$\,Dor \citep{kaye1999} and SPB stars \citep{waelkens1991} are main-sequence stars, with masses $1.4\,\mathrm{M}_\odot \lesssim M \lesssim 1.9\,\mathrm{M}_\odot$ and $3\,\mathrm{M}_\odot \lesssim M \lesssim 9\,\mathrm{M}_\odot$, respectively. Their g-mode pulsations have periods between 0.3 and 5 days, and are mostly sensitive to the near-core regions of the stars. Asteroseismic modelling of observed and identified g-mode pulsations allows us to constrain the physical processes taking place in the deep stellar interior \citep[see e.g.][]{aerts2018}. While similar on some fronts, major differences occur in the modelling of g~modes compared to the case of stochastically-excited pressure modes \citep[see][for an extensive review on the overall methodology]{aerts2020}. For g~modes in $\gamma$\,Dor and SPB stars, key processes to infer are convective core overshooting or convective penetration in the core boundary layers \citep[e.g.,][]{pedersen2018,michielsen2019}, interior magnetic fields \citep[][Bugnet et al. submitted]{prat2019,prat2020,vanbeeck2020, mathis2020}, and microscopic or macroscopic mixing in the radiative envelope \citep[e.g.,][Pedersen et al., submitted; Bugnet et al., submitted]{deal2016, rogers2017, pedersen2018, mombarg2020, mathis2020}.

One of the most crucial aspects that have to be taken into account in stellar structure and evolution theory is rotation \citep[][and references therein]{zahn1992, maeder+zahn1998, mathis+zahn2004, maeder2009}. Aside from causing numerous physical processes such as rotational mixing, stellar rotation severely influences the behaviour of g-mode pulsations via the Coriolis acceleration \citep[e.g.,][for a review]{lee+saio1997, dintrans+rieutord2000, aerts2019b}. For high-frequency oscillations ($\omega \gg \Omega$\,, with $\omega$ the angular pulsation frequency in the co-rotating frame and $\Omega$ the angular rotation rate) the Coriolis acceleration can be treated as a perturbation \citep{hansen1977, gough1981}. However, high-order g modes (with radial order $n$ $\gg$ spherical degree $\ell$) typically lie in the low-frequency range ($\omega \lesssim \Omega$), where the Coriolis force also contributes to the restoring of the oscillations. Such g-mode pulsations therefore occur in the gravito-inertial regime and correspond to gravito-inertial waves (GIW, hereafter). Hence, the Coriolis force can no longer be treated as a perturbation, and the hydrodynamical equations that govern the oscillations become an infinite set of coupled differential equations \citep{mathis2009}. However, approximate numerical solutions and general properties of GIW can still be obtained by truncating the infinite set of coupled differential equations \citep[e.g.,][]{berthomieu1978,lee+saio1986,lee+saio1987,dziembowski1987a,dziembowski1987b,dziembowski1987, dintrans1999, dintrans+rieutord2000, mathis2009}.

The influence of the Coriolis force on GIWs is commonly described using the traditional approximation of rotation (TAR). It was first developed by \citet{eckart1960} in his study of the dynamics of shallow atmospheres and oceans on Earth, and later introduced in stellar pulsation theory by \citet{berthomieu1978} and \citet{lee+saio1987}. The main assumption of the TAR is that the stratification in which the waves propagate, is sufficiently strong as to limit vertical wave motions. As a consequence, the horizontal component of the rotation vector and therefore also the vertical component of the Coriolis acceleration, can be neglected within the description of the GIWs. This condition for stable stratification (both in chemical composition and in entropy) is typically met in the radiative near-core region of $\gamma$\,Dor and SPB stars. By applying the TAR, the hydrodynamical oscillation equations can be  decoupled and rewritten in the form of the Laplace tidal equation \citep{laplace1799}.

Two additional assumptions made within the TAR are those of uniform rotation and spherical symmetry. \citet{mathis2009} abandoned the first assumption and included the effect of differential rotation within the framework of the TAR. Subsequently, the sensitivity of GIWs to the effect of differential rotation was assessed by \citet{vanreeth2018}. The assumption of spherical symmetry is valid when the star is rotating sufficiently slowly to ignore the centrifugal acceleration, i.e., $\Omega \ll \Omega_{\mathrm{c}}$, where $\Omega_{\mathrm{c}} = \sqrt{GM_{\star}/R_{\rm eq}^3} = \sqrt{8GM_{\star}/27R_{\rm pole}^3}\simeq \sqrt{8GM_{\star}/27R^3} $ is the Roche critical rotation rate, $G$ the universal gravitational constant and where
$R_{\rm eq}$ and $R_{\rm pole}$ stand for the equatorial and polar radius of the star, respectively \citep[see][Chapter\,2]{maeder2009}. However, a significant fraction of the $\gamma$\,Dor and SPB stars are moderate to fast rotators \citep[][Pedersen et al., submitted]{papics2017,li2020}. Hence, the effect of the centrifugal deformation should be taken into account in the theoretical description of their g-mode pulsations. 

\citet{mathis+prat2019} generalised the TAR for moderately-to-rapidly rotating stars by considering the effects of the centrifugal acceleration. First, they provide a prescription to deform a stellar structure model into a centrifugally deformed oblate spheroid. From the resulting perturbed physical quantities, a dimensionless deformation factor proportional to the square of the rotation rate can then be calculated. Secondly, this deformation factor is used to transition from a spherically symmetric to a spheroidal coordinate system, keeping only first-order terms in the deformation. Next, they re-derive the Laplace tidal equation within this new coordinate system, arriving at the so-called \emph{generalised} Laplace tidal equation. Finally, the authors derived an asymptotic expression for the frequencies of GIWs, including the effect of the centrifugal acceleration.

In this work, we have set up a parameter study with the goal of assessing the effect of the centrifugal acceleration on g-mode pulsations in rotationally deformed stars and their detectability in space-based photometric observations. By doing so, we expand upon the proof-of-concept study conducted in the theoretical work of \citet{mathis+prat2019} and aim to answer the question whether the effect of the centrifugal acceleration should be accounted for in asteroseimic modelling of observed g-mode pulsators.
A brief summary of the theoretical results and numerical implementation by \citet{mathis+prat2019}, as well as our improvements to the theoretical framework, are provided in Sect.\,\ref{sect:methodology}. The results of our upgraded implementation and parameter study follows in Sect.\,\ref{sect:results}. We discuss the results and conclude in Sect.\,\ref{sect:discussion}.

%%%%%%%%%%%%%%%%%%%%%%%%%%%%%%%%% METHODOLOGY %%%%%%%%%%%%%%%%%%%%%%%%%%%%%%%%%
\section{Methodology}\label{sect:methodology}
\subsection{Theoretical background}
\subsubsection{Deformation of stellar structure}\label{sect:deformation_stellar_structure}
Following \citet[][Appendix A]{mathis+prat2019}, the pressure $P$, density $\rho$ and gravitational potential $\phi$ in a centrifugally deformed star are written as the sum of a spherically symmetric, non-perturbed part (subscript `0') and a perturbation (subscript `1'). This perturbation term itself is then expanded on an orthogonal basis of Legendre polynomials $P_{\mathrm{leg},l}(\cos\theta)$ of degree $l=0,\,2$, with $\theta$ the co-latitude and $\theta=0$ on the rotation axis. Such an expansion corresponds to the projection of the perturbed quantities on a spheroidal surface in the moderately rapid rotation regime considered here. For fast rotators, additional terms of (even) degree $l$ should be included but that is beyond our current scope. 

For the gravitational potential, with $\phi_{0} = -GM(r)/r$ and $M(r)$ the mass contained within a sphere of radius $r$\,, we have:
\begin{equation}\label{eq:gravitational_potential_expansion}
    \phi(r, \theta) = \phi_{0}(r) + \phi_{1}(r, \theta) = \phi_{0}(r) + \sum_{l=0,2}\phi_{l}(r)P_{\mathrm{leg},l}(\cos\theta)\,. 
\end{equation}
With the inclusion of the centrifugal acceleration, the hydrostatic equilibrium in the star is written as:
\begin{equation}\label{eq:hydrostatic_equilibrium}
    \frac{\vec{\nabla}P}{\rho} = -\vec{\nabla}\phi + \frac{1}{2}\Omega^2 \vec{\nabla} \left(r^2 \sin^2\theta\right)\,. 
\end{equation}
This last term corresponds to the gradient of the centrifugal potential (per unit mass) $U(r,\theta) = -\frac{1}{2}\Omega^2r^2\sin^2\theta$. The centrifugal potential itself is then also expanded on the same basis of Legendre polynomials as:
\begin{equation}
    U = \sum_{l=0,2}U_{l}(r)P_{\mathrm{leg},l}(\cos \theta)\,.
\end{equation}
This implies that the modal amplitudes of the centrifugal potential are given by $U_{l=0} = -(1/3)\Omega^2r^2$ and $U_{l=2} = (1/3)\Omega^2r^2$.

The modal amplitudes of the pressure and density, $P_{l}(r)$ and $\rho_{l}(r)$, are recovered from the property that in the centrifugally deformed star the equipotential surface $(\phi + U)$, isobar surface and isodensity surface coincide and are given by \citep[][Appendix A]{mathis+prat2019}:
\begin{align}
    P_{l}(r) &= -\rho_{0}(r)\left[\phi_{l}(r) + U_{l}(r)\right] \label{eq:modal_amplitude_pressure}\,, \\
    \rho_{l}(r) &= \frac{1}{g_{0}(r)}\frac{\rm{d}\rho_{0}(r)}{\rm{d}r}\left[\phi_{l}(r) + U_{l}(r)\right] \label{eq:modal_amplitude_density}\,.
\end{align}
For the modal amplitude of the gravitational potential, $\phi_{l}(r)$\,, it is required to solve the \emph{perturbed} Poisson equation 
\begin{equation}\label{eq:perturbed_Poisson_equation}
    \nabla^2\phi_{l}(r) = 4\pi G\rho_{l}(r)\,,
\end{equation} which is simply recovered from substituting the modal expansions of $\phi(r,\theta)$ and $\rho(r,\theta)$ in the Poisson equation. Explicitly the perturbed Poisson equation reads:
\begin{equation}
    \frac{1}{r}\frac{\mathrm{d}^2}{\mathrm{d}r^2}\left(r \phi_{l}\right) - \frac{l\left(l+1\right)}{r^2}\phi_{l} - \frac{4\pi G}{g_{0}}\frac{\mathrm{d}\rho_{0}}{\mathrm{d}r}\phi_{l} = \frac{4\pi G}{g_{0}}\frac{\mathrm{d}\rho_{0}}{\mathrm{d}r}U_{l}\,,
\end{equation}
with boundary conditions
\begin{equation}
    \phi_{l}(0) = 0 \quad \mathrm{and} \quad \frac{\mathrm{d}}{\mathrm{d}r}\phi_{l}(R) = \frac{\left(l+1\right)}{R}\phi_{l}(R) 
\end{equation}
and $l \in \{0,\,2\}$ \citep[][Appendix A]{sweet1950,zahn1966, mathis+prat2019}.

Based on \citet{lee+baraffe1995} and \citet{mathis+prat2019}, we now define the pseudo-radial coordinate $a$ as follows:
\begin{equation}\label{r_ifo_a}
    r = a\left[1+\varepsilon(a,\theta)\right]\,.
\end{equation}
At this point, we perform a transformation from a spherical coordinate system $(r,\theta,\varphi)$ to a spheroidal coordinate system $(a,\theta,\varphi)$. $\varepsilon$ represents a dimensionless deformation factor, defined in the spheroidal coordinate system. Again following \citet[][Appendix A]{mathis+prat2019}, an expression for the modal amplitudes of $\varepsilon$ can be derived:
\begin{equation}\label{eq:modal_amplitude_epsilon}
    \varepsilon_{l}(a) = -\frac{\phi_{l}(A_{\mathrm{S}}) + U_{l}(A_{\mathrm{S}})}{g_{0}(A_{\mathrm{S}})}\frac{a^3}{A_{\mathrm{S}}^4} \,,
\end{equation}
with $A_{\rm S}$ the surface pseudo-radius of the deformed star. The coordinate mapping described by these expressions for $\varepsilon$ and $a$ is more accurate than the linear approximation used by \citet{mathis+prat2019} and closely agrees with the physical mapping described by, e.g., \citet{zahn1966} and \citet{mathis+zahn2004}, so that $r$ is equal to the deformed surface radius $R_{\rm S}(\theta)$ at $a=A_{\mathrm{S}}$. However, while this physical mapping has a singularity at the stellar centre, our cubic expression ensures that $a \simeq r$ when $a \rightarrow 0$.

From the expressions for the modal amplitudes of the pressure, density and deformation factor (Eqs.\,\ref{eq:modal_amplitude_pressure}, \ref{eq:modal_amplitude_density} \& \ref{eq:modal_amplitude_epsilon}), it can be seen that the centrifugal acceleration enters the expressions of the perturbed quantities in the form of a second order perturbation in the stellar rotation rate $\Omega$. In the expression for $\varepsilon_{l}(a)$ given in Eq.(\ref{eq:modal_amplitude_epsilon}) we have a dominant contribution from $U_l$ (compared to $\phi_l$). Hence, $\varepsilon_{l=0}$ ($\varepsilon_{l=2}$) is positive (negative), and the deformation factor $\varepsilon$ is positive in the direction of the equator.

To facilitate the calculation of $\varepsilon_l$, we can map the radial coordinate $r_0$ and the surface radius $R$ of the spherically symmetric (non-deformed) stellar model onto $a$ and $A_{\rm S}$, respectively. Within their respective stellar models, both the $r_0$ and $a$ coordinates coincide with the radial isobaric coordinate. Thus, we get
\begin{equation}\label{eq:map_radco}
    \varepsilon_{l}(r_0) = -\frac{\phi_{l}(R) + U_{l}(R)}{g_{0}(R)}\frac{r_0^3}{R^4} \,.
\end{equation}

\subsubsection{Generalised Laplace tidal equation}\label{sect:glte}
The Laplace tidal equation is an eigenvalue equation that can be derived from the oscillation equations by using the TAR. Its simple form immediately demonstrates the simplifying power of the latter approximation. Following \citet{lee+saio1997}, it can be written as:
\begin{equation}
    \mathcal{L}^{\rm{class.}}_{\nu m}\left[\Theta_{\nu km}\right] = -\Lambda_{\nu km}^{\mathrm{class.}}\Theta_{\nu km},
\end{equation}
with the operator $\mathcal{L}^{\rm{class.}}_{\nu m}$ defined as:
\begin{align}
    \begin{split}
    \mathcal{L}^{\rm{class.}}_{\nu m} =& \frac{1-x^2}{1-\nu^2 x^2}\partial^2_{x} - \frac{2x\left(1-\nu^2\right)}{\left(1-\nu^2 x^2\right)^2}\partial_{x} \\
    & \quad+ \left[\frac{m\nu\left(1+\nu^2 x^2\right)}{\left(1-\nu^2 x^2\right)^2} - \frac{m^2}{\left(1-x^2\right)\left(1-\nu^2 x^2\right)}\right]
    \end{split}\,.
\end{align}
In order to avoid confusion later on, we will refer to this equation as the \emph{classical} Laplace tidal equation (abbreviated as CLTE). Here, $\partial_x = \partial/\partial x$\,, $x = \cos\theta$\,, and $\nu = 2\Omega/\omega$ the spin parameter. The latter is a measure of the effect of rotation; pulsation modes for which $\nu > 1$ (resp. $\nu < 1)$ are known as sub- (resp. super-) inertial modes. The eigenfunctions $\Theta_{\nu km}$ are the radial Hough functions, after S.\,S. Hough, who pioneered in solving the Laplace tidal equation \citep{hough1898}, and $\Lambda_{\nu km}^{\mathrm{class.}}$ represent the eigenvalues of the equation. These radial Hough functions give the co-latitudinal distribution of the radial displacement of the star caused by the pulsations and in the limit $\nu \rightarrow 0$ reduce to $CP_l^{m}(x)$, with $C$ a constant and $P_{l}^{m}(x)$ the associated Legendre polynomial of degree $l$ and order $m$. The eigenvalues reduce to $\ell(\ell+1)$ without rotation.

The convention is adopted in which, considering $\nu > 0$\,, positive azimuthal orders $(m>0)$ denote \emph{prograde} modes and $m<0$ \emph{retrograde} modes. Since in general for each pair $(m,\nu)$ the CLTE yields an infinite set of solutions, the ordering number $k$ is introduced. Inertial waves, such as \emph{r} modes (which are normal modes of global Rossby waves influenced by buoyancy) \citep{saio2018}, are not present in non-rotating stars and have $k<0$. For $k\geq 0$, the ordering number is related to the spherical degree $l$ and azimuthal order through $\ell=|m| + k$ \citep{mathis+prat2019}.

To include the effect of the centrifugal acceleration in the Laplace tidal equation, it is re-derived in the spheroidal coordinate system by \citet{mathis+prat2019}. The transition from a spherical to spheroidal coordinate system involves the following basis transformation:
\begin{equation}\label{eq:basis_transformation}
 \begin{cases}
    \vec{\widetilde{e}}_{a} &= \left(1+\varepsilon+a\partial_{a}\varepsilon\right)\vec{\widehat{e}}_{r}, \\
    \vec{\widetilde{e}}_{\theta} &= \partial_{\theta}\varepsilon\vec{\widehat{e}}_{r} + \left(1+\varepsilon\right)\vec{\widehat{e}}_{\theta}, \\
    \vec{\widetilde{e}}_{\varphi} &= \left(1+\varepsilon\right)\vec{\widehat{e}}_{\varphi},
\end{cases}
\end{equation}
with $\varepsilon \equiv \varepsilon(a,\theta) = \sum_{l=0,2}\varepsilon_{l}(a)P_{\mathrm{leg},l}(x)$ the deformation factor introduced above. The resulting generalised Laplace tidal equation (GLTE) then is:
\begin{align}
\begin{split}
\mathcal{L}_{\nu m}\left[w_{\nu k m}\right]  =& \left(\frac{1-x^2}{\mathcal{D}}\right)\partial_x^2 w_{\nu km}\\ 
& + \left[\frac{\left(1-x^2\right)\partial_x\mathcal{E}}{\mathcal{D}} + \partial_{x}\left(\frac{1-x^2}{\mathcal{D}}\right)\right]\partial_x w_{\nu km} \\
& - \left[\frac{m^2}{\left(1-x^2\right)\mathcal{D}} - m\nu\frac{\mathrm{d}}{\mathrm{d}x}\left(\frac{x\mathcal{C}}{\mathcal{D}}\right) - m\nu\frac{x\mathcal{C}}{\mathcal{D}}\partial_{x}\mathcal{E}\right]w_{\nu km} \\
=&  -\Lambda_{\nu km}(a)w_{\nu km}
\end{split}\,,
\end{align}
where
\begin{align*}
    \mathcal{A}(a,\theta) &= 1+2\varepsilon\,,\quad \mathcal{B}(a,\theta) = \mathcal{A} + \tan\theta\partial_{\theta}\varepsilon \\
    \mathcal{C}(a,\theta) &= \mathcal{B}/\mathcal{A}\,, \quad \mathcal{D}(a,\theta) = \mathcal{A}\left[1-\nu^2\cos^2\theta\,\mathcal{C}^2\right] \\
    \mathcal{E}(a,\theta) &= 3\varepsilon + a\partial_{a}\varepsilon\,.
\end{align*}
Utilising the same nomenclature as \citet{mathis+prat2019}, the eigenfunction $w_{\nu km} = w_{\nu km}(a,\theta)$ is called the modified radial Hough function. Note that the eigenvalue $\Lambda_{\nu km}(a)$ and modified radial Hough function have a parametric dependence on the pseudo-radial coordinate $a$, whereas in the spherically symmetric case there was no radial dependence of the solutions. 

\subsubsection{Asymptotic frequencies}\label{sect:asymptotic}
Within the asymptotic regime, where the radial order $n$ of the modes is much larger than the spherical degree $\ell$ ($n \gg \ell$), and where we consider modes with $\omega \ll |N|$ and $\omega \ll |S_{\ell}|$, $N$ and $S_{\ell}$ being the Brunt-V\"ais\"al\"a and Lamb frequency respectively, asymptotic expressions for the pulsation frequencies can be derived \citep{shibahashi1979, tassoul1980, unno1989}. Following \citet{mathis2009} and \citet{bouabid2013}, these asymptotic expressions for the pulsation frequencies can be re-derived to include the effect of the centrifugal force. \citet{mathis+prat2019} find 
the angular pulsation frequencies to be:
\begin{equation}\label{eq:asymptotic_frequencies}
    \omega_{nkm} = \frac{\displaystyle{\int_{a_{1}}^{a_{2}}} \frac{\Lambda_{\nu km}^{1/2}(a)\bar{N}(a)}{a}\mathrm{d} a}{(n+1/2)\pi}
\end{equation}
and the pulsation periods:
\begin{equation}
\label{Pnkm}
P_{nkm} = \frac{2\pi^2(n+1/2)}{\displaystyle{\int_{a_{1}}^{a_{2}} \frac{\Lambda_{\nu km}^{1/2}(a)\bar{N}(a)}{a}\mathrm{d} a}}\,.
\end{equation}
Here, $\bar{N}(a)$ is the perturbed Brunt-V\"ais\"al\"a frequency profile (quantities denoted with a bar are defined within the spheroidal coordinate system). To account for the large diversity of pulsation mode cavities of \emph{g}~modes with different identification $(k,m)$ or spin $\nu$, $\bar{N}^2(a)$ is calculated as a weighted average over the co-latitude $\theta$:
\begin{equation}
    \bar{N}^2(a) = \frac{\int_0^\pi H_r(a,\theta) N^2(a,\theta)\sin\theta\,\mathrm{d}\theta}{\int_0^\pi H_r(a,\theta)\sin\theta\,\mathrm{d}\theta}\,,
\end{equation}
where $H_r(a,\theta)$ is the radial eigenfunction that corresponds to the eigenvalue $\Lambda_{\nu km}(a)$ in Eqs.(\ref{eq:asymptotic_frequencies}) and (\ref{Pnkm}), and
\begin{equation}
N^2(a,\theta) = -\frac{\bar{g}}{r}\left[\frac{\mathrm{d}\ln\bar{\rho}}{\mathrm{d}\ln r} - \frac{1}{\bar{\Gamma}_{1}}\frac{\mathrm{d}\ln \bar{P}}{\mathrm{d}\ln r}\right] \,,\label{eq:N2bar_2d}
\end{equation}
with $\bar{\Gamma}_{1} = \left(\partial \ln\bar{P}/\partial \ln\bar{\rho}\right)_{\bar{S}}$ the perturbed adiabatic exponent defined at constant (perturbed) entropy $\bar{S}$. The different quantities on the right-hand side of Eq.(\ref{eq:N2bar_2d}) all depend on both $a$ and $\theta$. In the integral in Eq.\,(\ref{Pnkm}), $a_{1}$ and $a_{2}$ are the inner- and outer-boundaries of the mode cavity, determined as the region(s) within the star where $\bar{N}^2(a) > 0$.

Another result from asymptotic theory is that, for a non-rotating, non-magnetic g-mode pulsator without chemical gradients, the pulsation periods of modes with consecutive radial orders are equidistant. The differences in mode periods, $\Delta P_{nkm} = P_{n+1,km} - P_{nkm}$ thus remain constant when plotted as a function of the period $P_{nkm}$ or radial order $n$ \citep{miglio2008}. When constructing these so-called period spacing patterns for more realistic stellar models (including rotation and chemical gradients), they will deviate from a constant, as first observed for a main-sequence massive early-type star by the CoRoT satellite \citep{Degroote2010}. Comparing theoretical and observed period spacing patterns then allows us to deduce information about rotation and chemical mixing in the deep interior of stars \citep{aerts2020}.

\subsection{Computational implementation}
The computational method to derive the asymptotic GIW frequencies from the solutions to Eq.\,(\ref{eq:asymptotic_frequencies}) is provided in Appendix\,\ref{app:asymptotic}. The implementation of the analytical framework of \citet{mathis+prat2019} described above consists of three parts. First, a non-rotating 1D stellar structure model is calculated and perturbed for a selected uniform rotation rate, following Sect.\,\ref{sect:deformation_stellar_structure}. These 1D, non-rotating stellar structure models are calculated using the code \texttt{MESA}\footnote{\texttt{MESA} version r11701. For more information about the different versions, see \url{http://mesa.sourceforge.net/}.} (Modules for Experiments in Stellar Astrophysics; \citealt{paxton2019} and references therein). The resulting 2D (radial and latitudinal) profiles for $r(a,\theta)$ and $\varepsilon$ are subsequently used as input for solving the GLTE given in Sect.\,\ref{sect:glte}. The solutions of the GLTE then allow us to compute the asymptotic pulsation frequencies based on Eq.\,(\ref{eq:asymptotic_frequencies}). 

The computational implementation of the first step involves solving the perturbed Poisson equation (e.g.\ through an iterative shooting scheme) for $l=0$ and $l=2$, yielding the modal amplitudes $\phi_{l}(r)$. These, in turn, are used to calculate the modal amplitudes of the pressure and density (Eqs.\,\ref{eq:modal_amplitude_pressure}\,\&\,\ref{eq:modal_amplitude_density}). 
Two challenges occur in the computation of the perturbed Brunt-V\"ais\"al\"a frequency profile:
\begin{enumerate}[i]
    \item The calculation of $\bar{\Gamma}_{1}$ (at constant entropy $\bar{S}$) requires detailed knowledge on how the equation of state changes under influence of the centrifugal deformation. This would require the framework of \citet{mathis+prat2019} to be directly implemented in \texttt{MESA}.
    \item High levels of numerical noise are introduced by the numerical derivatives of the physical quantities (such as the pressure and density) with respect to $r$ and $a$, which is a well-known problem in stellar structure and evolution codes \citep{paxton2013}.
\end{enumerate}
The details of how we treated these aspects are described in Appendix\,\ref{app:BV_frequency}.

\subsubsection{Solver for the GLTE}\label{sect:solver_for_GLTE}
Several numerical methods are available for solving an eigenvalue problem such as the Laplace tidal equation, a number of which are discussed in \citet{wang2016}. As in \citet{mathis+prat2019}, the Chebyshev collocation method \citep{boyd1976} is chosen. In such a method, the eigenfunctions of the problem, in this case the modified radial Hough functions, are expanded on a basis of $N$ Chebyshev polynomials, where $N$ is the number of collocation points. For the current work, we improved an existing solver for the CLTE of \citet{prat2019} and \citet{vanbeeck2020}, which was developed by one of us (VP), with two modifications.

Contrary to the CLTE, the GLTE has a parametric dependence on the radial coordinate (in this case $a$). In other words, at each cell of the \texttt{MESA} model, the solutions of the GLTE will be different, since $\varepsilon(a,\theta)$ depends on $a$. Furthermore, in order to select the solution corresponding to the desired mode identification $(k,m)$ among the set of $N$ solutions, the solver requires an estimate for the eigenvalue $\Lambda_{\nu km}(a)$. At the centre of the model, where $\varepsilon = 0$, the eigenvalue can be estimated by the solutions of the CLTE tabulated in the TAR module of the stellar pulsation code \texttt{GYRE} \citep{townsend2013,townsend2018}. Going from the centre to the surface, at each subsequent cell, the estimate is then provided by the eigenvalue from the previous cell. Since the effect of the centrifugal force increases from the centre to the surface (see Eqs.\,\ref{eq:hydrostatic_equilibrium}\,\&\ref{eq:modal_amplitude_epsilon}) these eigenvalues will increasingly diverge from the value at the centre.
To keep the computation time manageable, an appropriate number of sample points in which  the GLTE is solved, must be selected. Thanks to the smooth behaviour of $\Lambda_{\nu km}(a)$ (see Sect.\ref{sect:results}) it suffices to calculate solutions for $\sim 10$ cells. These points are chosen equidistant in cell index in the \texttt{MESA} model rather than in physical distance. In that way, regions with higher cell density get a higher sampling. $\Lambda_{\nu km}(a)$ profiles, required for asymptotic frequency calculations (see Eq.\,\ref{eq:asymptotic_frequencies}) are retrieved through quadratic interpolation.

The calculation of the coefficients $\mathcal{A}, \mathcal{B}, \mathcal{C}, \mathcal{D}, \mathcal{E}$ and their derivatives, which make up the coefficients of the GLTE, is subject to discontinuities when done from numerical differentiation. We therefore relied on analytical differentiation where possible. Only for $\mathcal{E}$ and its derivative with respect to the co-latitudinal coordinate $x$ we approximate the term $a\partial_{a}\varepsilon$ by $3\varepsilon$ in order to avoid numerical issues. This approximation is justified by the cubic polynomial behaviour of the deformation factor. The coefficients, as implemented in the solver for the GLTE, are:
\begin{align}
    \partial_{x}\varepsilon &= 3x\varepsilon_{l=2}\label{eq:dxvareps}\\
    \mathcal{A} &= 1+2\varepsilon\\
    \mathcal{B} &= \mathcal{A} - 3\varepsilon_{l=2}(1-x^2)\\
    \mathcal{C} &= \mathcal{B}/\mathcal{A}\\
    \mathcal{D} &= \mathcal{A}\left(1-\nu^2x^2\mathcal{C}^2\right)\\
    \mathcal{E} &\simeq 6\varepsilon\,,\label{eq:evar}
\end{align}
with $\varepsilon_{l=2} = \varepsilon_{l=2}(a)$ the modal amplitude in the expression for $\varepsilon = \varepsilon(a,\theta) = \sum_{l=0,2}\varepsilon_{l}(a)P_{\mathrm{leg},l}(x)$. The respective derivatives with respect to $x=\cos\theta$ are:
\begin{align}
    \partial_{x}\mathcal{A} &= 6x\varepsilon_{l=2}\\
    \partial_{x}\mathcal{B} &= 12x\varepsilon_{l=2}\\
    \partial_{x}\mathcal{C} &= \frac{1}{\mathcal{A}^2}\left(\mathcal{A}\partial_{x}\mathcal{B} - \mathcal{B}\partial_{x}\mathcal{A}\right)\\
    \partial_{x}\mathcal{D} &= \partial_{x}\mathcal{A}\left(1-\nu^2x^2\mathcal{C}^2\right) - 2\mathcal{A}\mathcal{C}x\nu^2\left(\mathcal{C} + \partial_{x}\mathcal{C}\right)\\
    \partial_{x}\mathcal{E} &= 18x\varepsilon_{l=2}\,.
\end{align}

%%%%%%%%%%%%%%%%%%%%%%%%%%%%%%%%% RESULTS %%%%%%%%%%%%%%%%%%%%%%%%%%%%%%%%%
\section{Numerical results}\label{sect:results}

\subsection{Covered parameter space of equilibrium models}
\begin{figure}
    \centering
    \resizebox{\hsize}{!}{\includegraphics{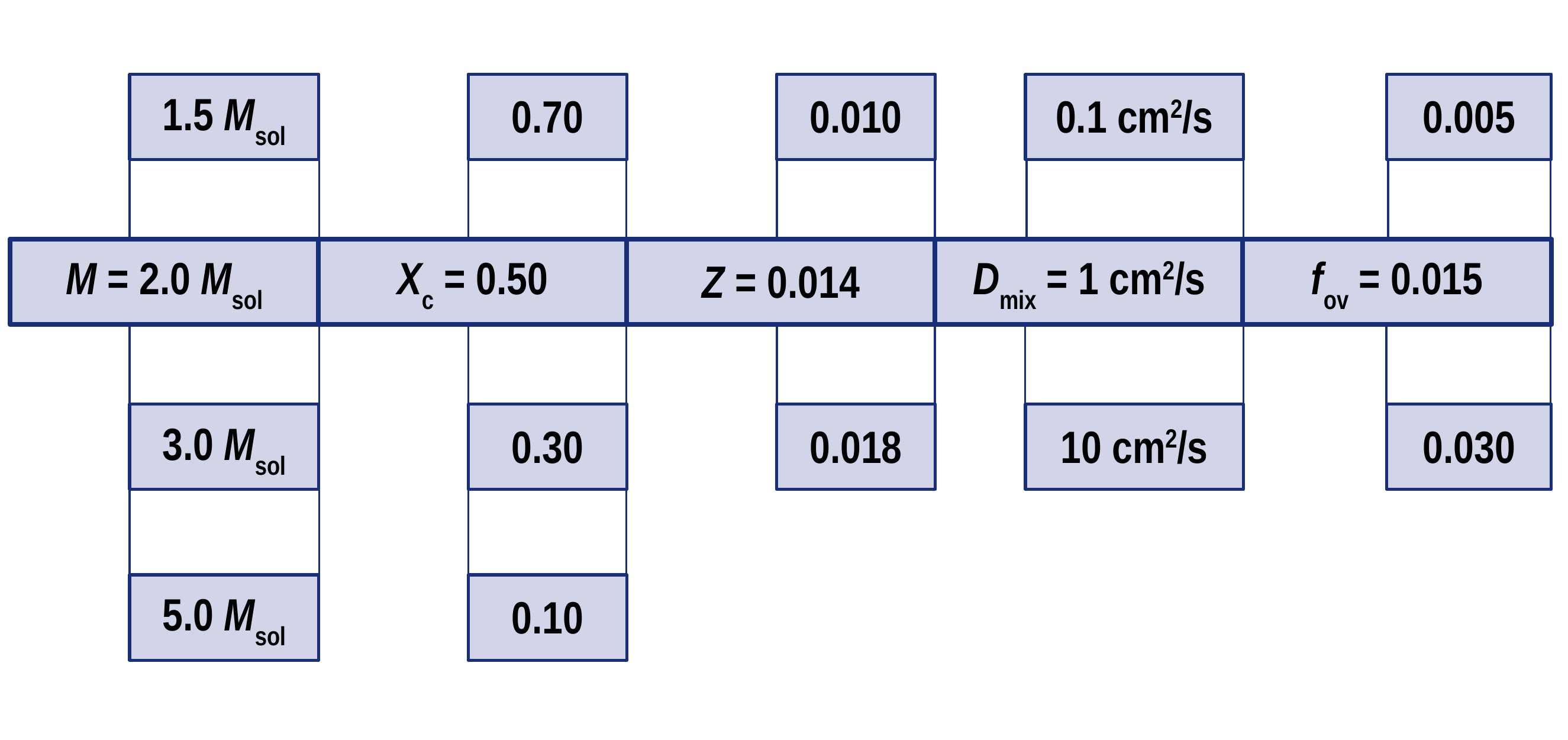}}
    \caption{Selected ranges and values for the input parameters of the 1D stellar structure models used in this work.}
    \label{fig:model_tree}
\end{figure}

With the goal of exploring the effect of the centrifugal acceleration on the stellar structure and on high-order g-mode pulsations in more depth than was done in the proof-of-concept by \citet{mathis+prat2019}, the methodology described in the previous section has been applied to a range of \texttt{MESA} equilibrium models with varying input parameters. The ranges and values for these parameters are shown in Fig.\,\ref{fig:model_tree}.

\begin{itemize}
\item \textbf{mass:} the masses of the \texttt{MESA} models are chosen within the joint mass range of $\gamma\,$Dor $[1.4,1.9\,\mathrm{M}_{\odot}]$ \citep{mombarg2019} and SPB $[3,9\,\mathrm{M}_{\odot}]$ 
\citep[][Pedersen et al., submitted]{papics2017} stars. The physical conditions inside these types of pulsating stars allow the TAR to be applied and these are the two types of main-sequence stars for which g-mode period spacing patterns are observed. 

\item \textbf{age:} the age of the stellar models is quantified in terms of their core hydrogen-mass fraction $X_{\rm{c}}$. Each model has been computed with an initial hydrogen mass fraction $X_{\mathrm{ini}} = 0.715$ and evolved to its specified $X_{\mathrm{c}}$.

\item \textbf{metallicity:} the range in initial metal mass fraction, $Z$, covers values typical for stars in our Milky Way in the considered mass range. By considering a range, we assess how the change in opacity caused by $Z$ affects the Brunt-V\"ais\"al\"a profile and the g-mode pulsation frequencies.

\item \textbf{envelope mixing:} the amount of mixing that occurs in the radiative envelope of the star is quantified by the diffusion coefficient $D_{\rm{mix}}$ and affects the profiles of the mass fractions, $X_i$, of all the isotopes considered in the chemical mixture adopted as input physics. As such, $D_{\rm mix}$ has an influence on the Brunt-V\"ais\"al\"a frequency. In each of the calculated \texttt{MESA} models, we considered $D_{\rm{mix}}$ to be constant throughout the radiative envelope, adopting values typical for g-mode pulsators \citep{vanreeth2016,moravveji2016}.

\item \textbf{convective core overshooting:} this process is of major importance for stellar evolution in the considered mass range, yet it is least known among the various ingredients to be chosen as input physics \citep{torresclaret2019,johnston2019,lisundegrijs2019,tkachenko2020}. Different formalisms exist to describe this overshoot region. We adopted a diffusive exponentially-decaying overshooting \citep{freytag1995,herwig2000} quantified by the overshooting parameter $f_{\rm{ov}}$. 
\end{itemize}

Furthermore, each model is computed using the AGSS09 chemical abundances derived by \citet{asplund2009} and a mixing length $\alpha_{\mathrm{MLT}} = 2$ within the mixing length theory developed by \citet{henyey1965}. For convergence purposes, a hot wind with Vink scaling factor of 1 is turned on \citet{vink2001}. In order to avoid numerical issues, rotation is not taken into account during the \texttt{MESA} model calculations. The effects of rotation enter in our models by applying the perturbation method described in Sect.\,\ref{sect:deformation_stellar_structure}.

\subsection{Solutions of the GLTE}
Here, we investigate the difference in solutions of the GLTE and CLTE. To illustrate these, we pick one baseline equilibrium model from our grid setup, as indicated in the highlighted row in Fig.\,\ref{fig:model_tree}. The behaviour is found to be equivalent for the other models so we do not discuss these results for brevity.

\begin{figure}
    \centering
    \resizebox{\hsize}{!}{\includegraphics{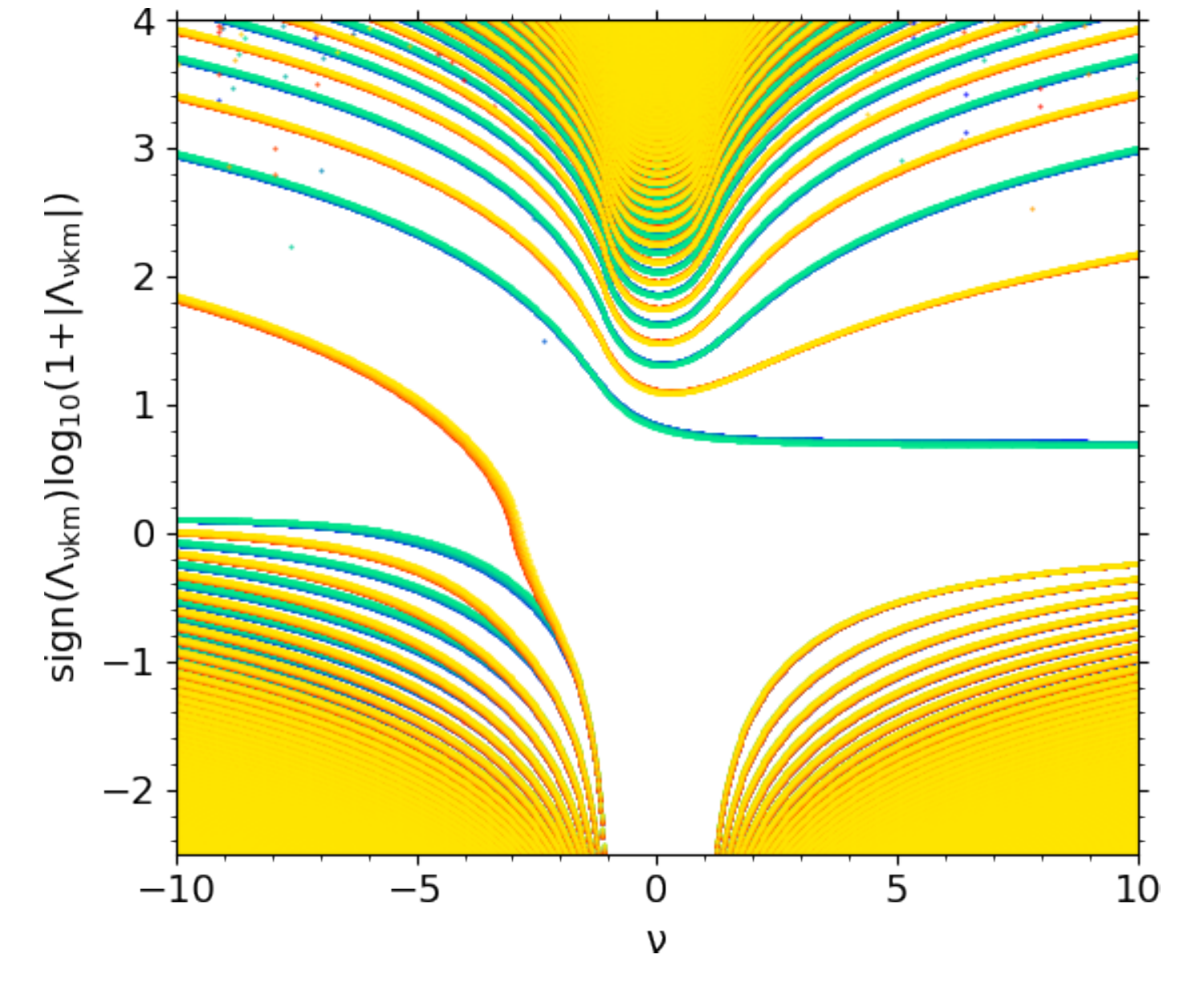}}
    \caption{Solution spectrum of the generalised Laplace tidal equation for modes with $m=2$. Modes with even eigenfunctions are shown in blue, those with odd eigenfunctions in orange. Light/dark colours correspond to solutions near the surface/core. The equilibrium model used as input for the computations has parameters as in the highlighted row in Fig.\,\ref{fig:model_tree}.}
    \label{fig:spectrum}
\end{figure}

Fig.\,\ref{fig:spectrum} shows the solution spectrum for the GLTE for $m=2$ with even and odd eigenfunctions. Four classes of solutions can be distinguished \citep{lee+saio1997}: (i) the prograde g modes with $\nu > 0$ and positive eigenvalues $\Lambda_{\nu km}(a)$, (ii) retrograde g modes with $\nu < 0$ and positive eigenvalues, (iii) Rossby modes which are retrograde and have negative eigenvalues, 
and (iv) prograde convective modes with negative eigenvalues. Just as Rossby modes, the latter only appear in rotating stars and only exist for $|\nu| > 1$. Specifically, these modes are able to propagate in convective regions under the joint force of the Coriolis acceleration and buoyancy.

Keeping in mind that the solutions plotted in the darkest colours correspond to those for the core and are therefore equivalent to the solutions of the CLTE (see Sect.\,\ref{sect:solver_for_GLTE}), Fig.\,\ref{fig:spectrum} reveals that the centrifugal deformation of the star causes a gradual shift in the eigenvalues. Whether this is an upwards or downwards shift depends on the mode identification $(k,m)$ (we recall that solutions for negative spin parameters $\nu$ and $m>0$ are equivalent to solution with $\nu > 0$ and $m<0$). The numerical `noise' visible at higher values of $|\nu|$ are an artefact due to the limited number of Chebyshev collocation points ($N=200$), but it can be seen that this does not affect the selected solutions. We obtained that an increase in the number of collocation points causes a decrease in this noise. 

\begin{figure}
    \centering
    \resizebox{\hsize}{!}{\includegraphics{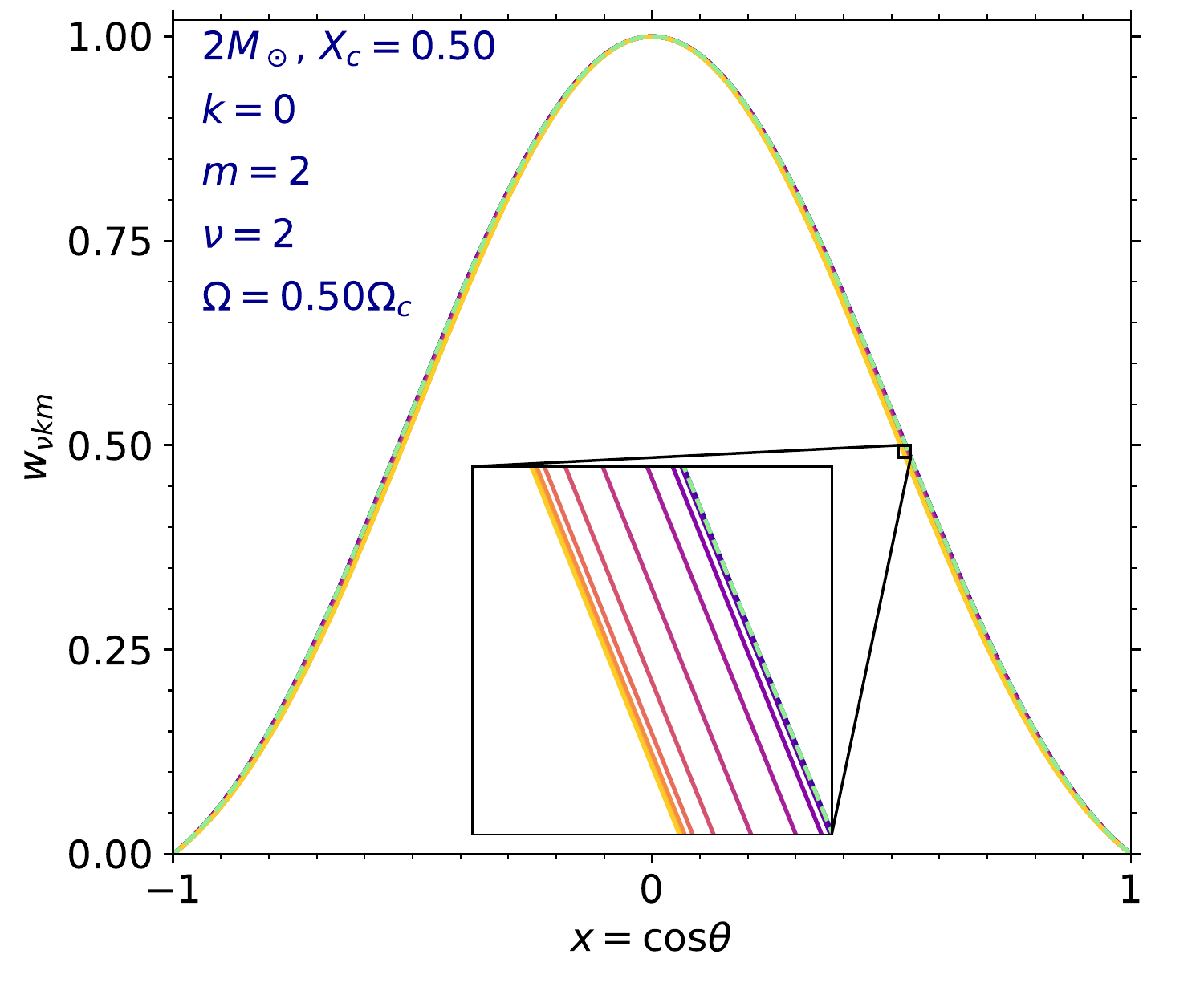}}
    \caption{Modified radial Hough functions $w_{\nu km}(a,\theta)$ (normalised) for $\nu=2$ and $\Omega/\Omega_{\mathrm{c}}=0.50$. Solutions are plotted in a colour range from indigo to yellow from the stellar core to the surface. The green dashed line shows the solution of the CLTE. The number of collocation points for this computation was $N=200$.}
    \label{fig:k0m2nu2Om30_hough}
\end{figure}

\begin{figure}
    \centering
    \resizebox{\hsize}{!}{\includegraphics{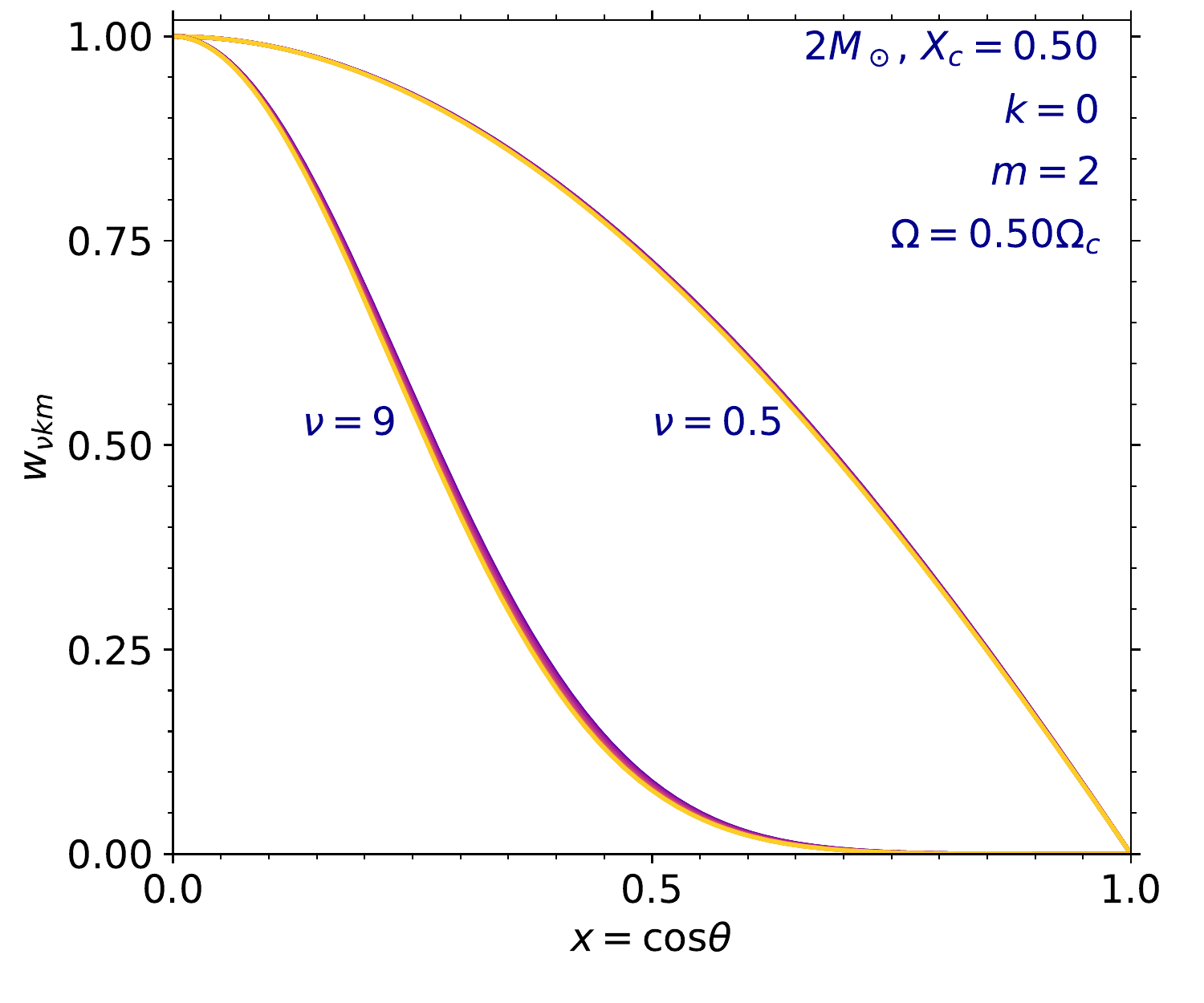}}
    \caption{Modified radial Hough functions $w_{\nu km}(a,\theta)$ (normalised) for $\Omega/\Omega_{\mathrm{c}}=0.50$ and two different spin parameters. The $2.0\,\mathrm{M}_{\odot}$, $X_{\rm{c}}=0.50$ equilibrium model was used as input. The colour scheme and number of collocation points are identical to that of Fig.\,\ref{fig:k0m2nu2Om30_hough} (CLTE solutions are left out for clarity).}
    \label{fig:nu05_9_houghs}
\end{figure}

\begin{figure*}
    \centering
    \includegraphics[width=\textwidth]{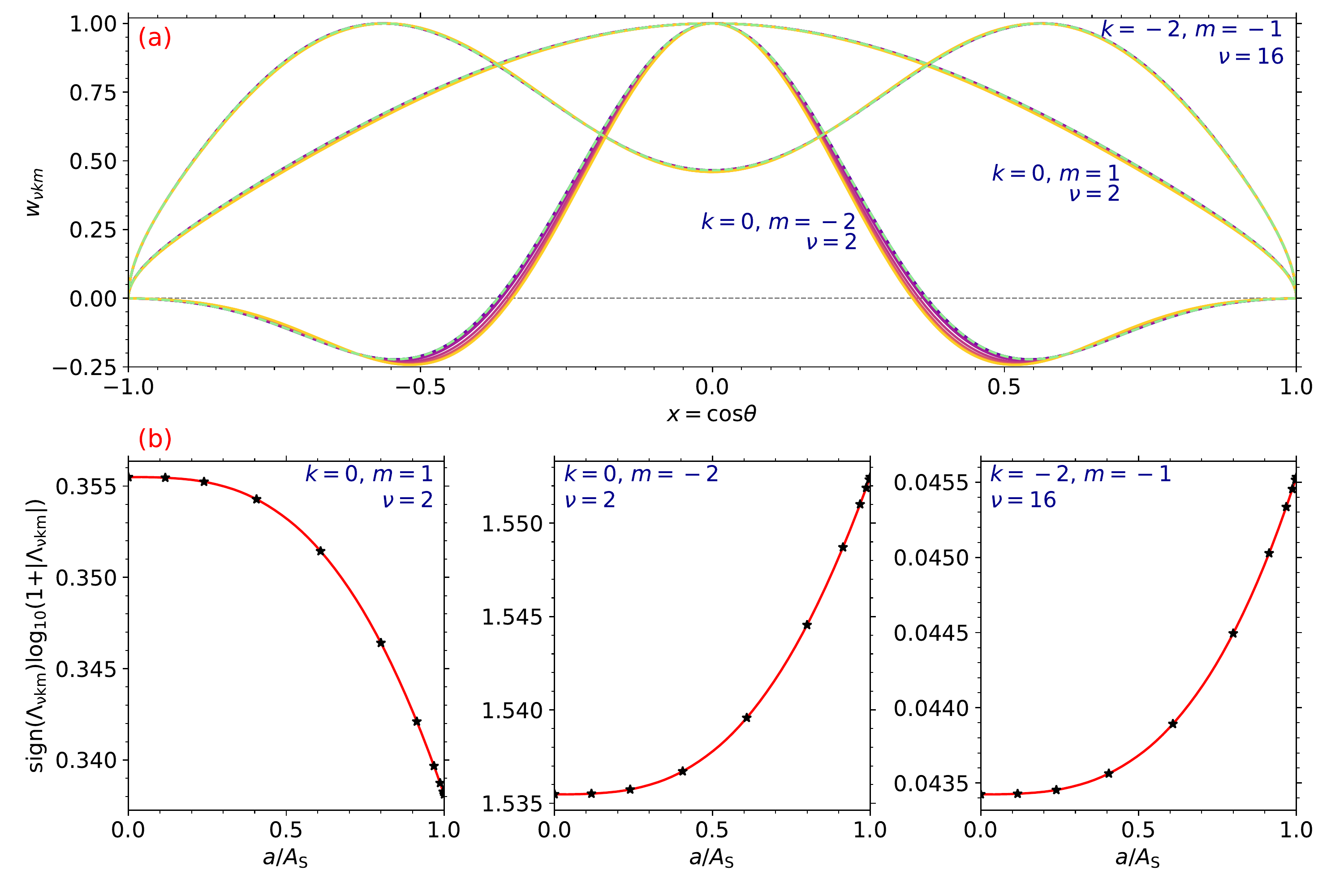}
    \caption{\emph{(a)} Modified radial Hough functions $w_{\nu km}(a,\theta)$ (normalised) for three different modes (with specified spin parameter $\nu$). Colour conventions are the same as those in Fig.\,\ref{fig:k0m2nu2Om30_hough}. \emph{(b)} Eigenvalue $\Lambda_{\nu km}(a)$ profiles corresponding to the eigenfunctions in \emph{(a)}. Points indicated with a $\star$ correspond to eigenvalues obtained by solving the GLTE. Red points are found via quadratic interpolation.
    The considered number of collocation points was $N=200$.}
    \label{fig:other_houghs}
\end{figure*}

The (normalised) eigenfunctions $w_{\nu km}$ of the GLTE for a $(k=0, m=2)$ mode with $\nu=2$ are shown in Fig.\,\ref{fig:k0m2nu2Om30_hough}. Equivalent figures for $\nu=0.5$ (super-inertial) and $\nu=9$ (sub-inertial) are displayed in Fig.\,\ref{fig:nu05_9_houghs}. The eigenfunctions of the GLTE differ increasingly from those of the CLTE as the distance from the centre of the model to the surface increases. More specifically, the eigenfunctions migrate inwards, towards the equator ($x = 0$), causing a narrowing of the overall shape of the eigenfunctions. Similar behaviour is observed for other modes, such as prograde dipole $(k=0,m=1)$ modes shown in Fig.\,\ref{fig:other_houghs}, retrograde quadrupole $(k=0, m=-2)$, retrograde Rossby $(k=-2,m=-1)$, quadrupole zonal $(k=2, m=0)$, and $(k=1,m=1)$ oscillation modes. Off-equator extrema, such as for the eigenfunctions of $(k=0,m=-2)$ and $(k=-2,m=-1)$ modes shown in Fig.\,\ref{fig:other_houghs}, 
experience net inward shifts towards the equator. 

\begin{figure}
    \centering
    \includegraphics[width=88mm]{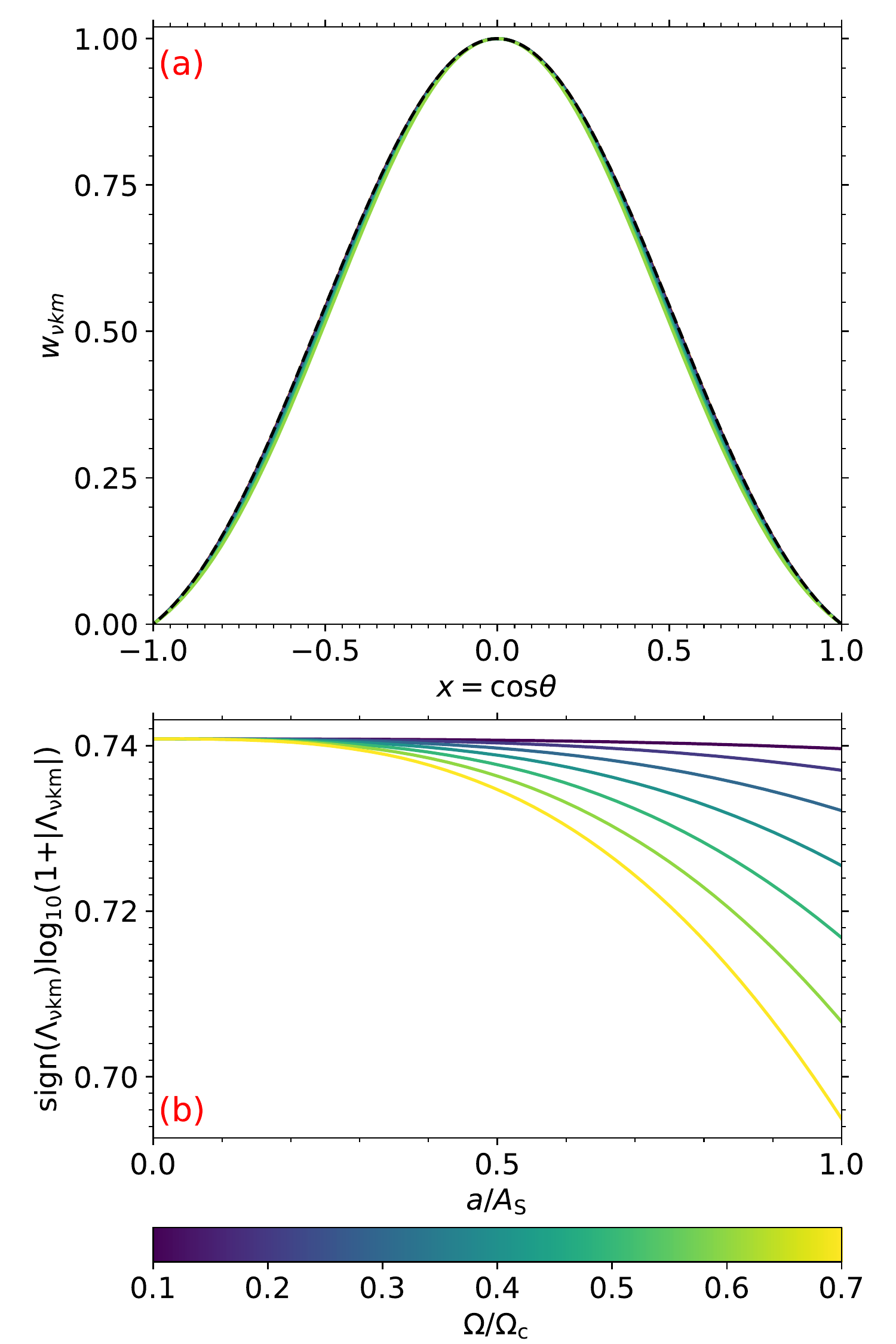}
    \caption{\emph{(a)} Modified radial Hough functions $w_{\nu km}(a=A_{\mathrm{S}},\theta)$ for seven different rotation rates and $\nu = 2$\,. \emph{(b)} Eigenvalue $\Lambda_{\nu km}(a)$ profiles corresponding to the eigenfunctions in \emph{(a)}\,.}
    \label{fig:hough_rotation}
\end{figure}

Fig.\,\ref{fig:hough_rotation} shows the eigenfunctions at the surface $(a=A_{\mathrm{S}})$ and the $\Lambda_{\nu km}(a)$ profiles for a $M=1.5\,\mathrm{M}_{\odot}$, $X_{\mathrm{c}}=0.50$ ($X_{\rm{c}}/X_{\rm{ini}}=0.70$) model for seven different rotation rates ranging from $\Omega = 0.1\,\Omega_{\rm{c}}$ to $\Omega = 0.7\,\Omega_{\rm{c}}$. The figure is again for a quadrupole sectoral $(k=0,m=2)$ mode. We find that the narrowing of the shape of the eigenfunctions and the divergence of the eigenvalues from the value at the centre increase with increasing rotation rate, as expected. 

\subsection{Asymptotic period spacing patterns}\label{sect:asymptotic_period_spacing_patterns}
\begin{figure}
    \centering
    \resizebox{\hsize}{!}{\includegraphics{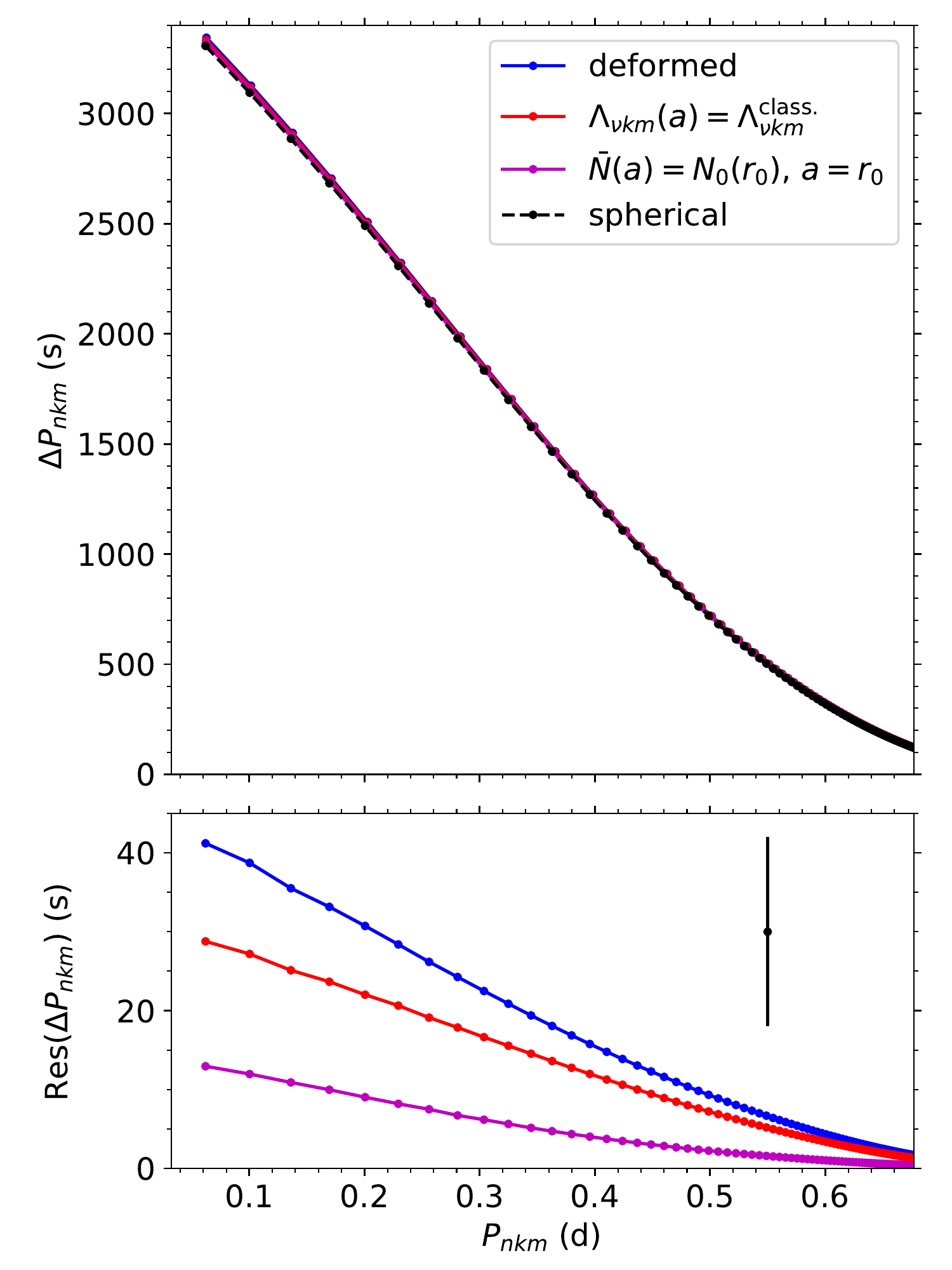}}
    \caption{Period spacing pattern in an inertial frame computed for $(k=0,m=1)$ modes in a centrifugally deformed $2.0\,\mathrm{M}_{\odot}$, $X_{\rm{c}}=0.50$ ($X_{\rm{c}}/X_{\rm{ini}}=0.70$), $Z=0.014$, $D_{\mathrm{mix}}=1\,\mathrm{cm}^2\mathrm{s}^{-1}$, $f_{\mathrm{ov}}=0.015$ equilibrium model at $\Omega/\Omega_{\mathrm{c}}=0.50$ (blue line). The black-dashed line shows the equivalent period spacing pattern for the same model, but without the effect of the centrifugal acceleration. The red and purple lines show the results for $N_{0}(r)\rightarrow\bar{N}(a);\,r\rightarrow a$ and $\Lambda_{\nu km}^{\mathrm{class.}}\rightarrow \Lambda_{\nu km}(a)$, respectively. The bottom panel shows the differences with respect to the spherically symmetric period spacing pattern. The indicated global errors for the periods and period spacings are averages calculated from the uncertainties of $P$ and $\Delta\,P$ 
for a sample of 40 $\gamma\,$Dor stars by \citet{vanreeth2015}, where the horizontal bar is hardly visible as it represents $10^{-4}\,$d.}
    \label{fig:psp_dipole_central}
\end{figure}

We solved the GLTE for each model in our grid, covering a range of spin parameters for each of the models (see Appendix\,\ref{app:asymptotic}). We restricted to three specific mode identifications: prograde dipole sectoral $(k=0,m=1)$ modes, prograde quadrupole sectoral $(k=0,m=2)$ modes, and retrograde Rossby modes with $(k=-2,m=-1)$. The reason for this choice is that these GIWs are most often observed in rotating g-mode pulsators \citep{papics2017,li2020}. The computed period spacing pattern for prograde dipole modes of the baseline model for $X_{\rm{c}}=0.50$ ($X_{\rm{c}}/X_{\rm{ini}}=0.70$) is shown in Fig.\,\ref{fig:psp_dipole_central}. By comparing the centrifugally deformed period spacing pattern with their spherically symmetric counterpart, we find that the spacing values increase under the influence of the centrifugal acceleration. This increase is largest at lowest radial orders (short pulsation periods). Similar behaviour is found for $(k=0,m=2)$ modes.

\begin{figure}
    \centering
    \resizebox{\hsize}{!}{\includegraphics{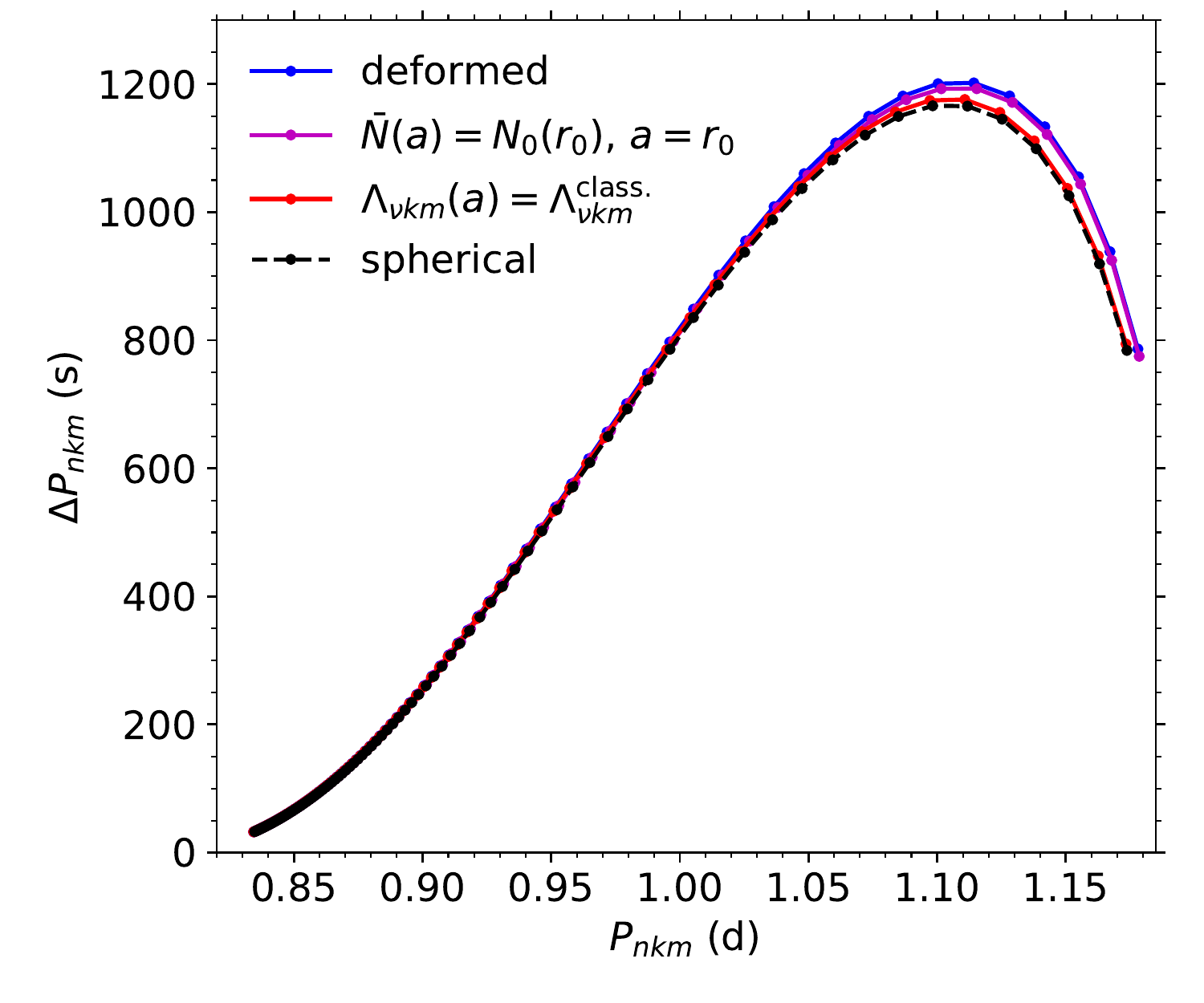}}
    \caption{Theoretical period spacing pattern in the inertial frame for $(k=-2,m=-1)$ modes in the same equilibrium model as in Fig.\,\ref{fig:psp_dipole_central} at $\Omega/\Omega_{\mathrm{c}}=0.50$ (blue line). The black-dashed line shows the equivalent period spacing pattern for the same model, but without the effect of the centrifugal acceleration. The red and purple show the isolated effect of $N_{0}(r)\rightarrow\bar{N}(a);\,r\rightarrow a$ and $\Lambda_{\nu km}^{\mathrm{class.}}\rightarrow \Lambda_{\nu km}(a)$, respectively.}
    \label{fig:psp_rossby_central}
\end{figure}

Essentially two effects are at play here: on the one hand, the centrifugal acceleration directly affects the g-mode pulsations, via a $\Lambda_{\nu km}(a)-$profile that changes throughout the star from $a=0$ to $a=A_{\mathrm{S}}$ instead of 
just one $\Lambda_{\nu km}^{\mathrm{class.}}$ value for $r_0=0\rightarrow r_0=R$. On the other hand, the stellar shape, including the Brunt-V\"ais\"al\"a frequency profile and the volume, are perturbed, which has an indirect on the \emph{g}-mode pulsations. In order to isolate the effect of this deformation, we re-calculated the period spacing pattern for the centrifugally deformed star for  $\Lambda_{\nu km}(a) = \Lambda_{\nu km}^{\mathrm{class.}}$, with $\Lambda_{\nu km}^{\mathrm{class.}}$ the eigenvalue of the CLTE. This period spacing pattern is plotted in red in Fig.\,\ref{fig:psp_dipole_central}. Similarly, the purple period spacing pattern was calculated by setting $\bar{N}(a) = N_{0}(r_0)$ and $a=r_0$, isolating the effect of the $a-$dependence of the solutions of the GLTE. Both effects lead to a net increase in the period spacings.
Also for the $(k=-2,m=-1)$ Rossby modes shown in Fig.\,\ref{fig:psp_rossby_central}, a net upward shift of the period spacing pattern occurs. The effect is larger at longer pulsation periods, which for Rossby modes is equivalent to low radial orders \citep{saio2018}. In some cases, such as for retrograde g~modes with $(k,m)$ = (0,-1), we do find a net decrease in the period spacings $\Delta P_{nkm}$. In such cases the $\Lambda_{\nu km}(a)-$value increases under the influence of the centrifugal acceleration, and counteracts the effect of the centrifugally modulated Brunt-V\"ais\"al\"a frequency profile $\bar{N}(a)$.

\subsection{Detectability in space-based photometric observations}\label{sect:detectability}
\begin{figure}
    \centering
    \resizebox{\hsize}{!}{\includegraphics{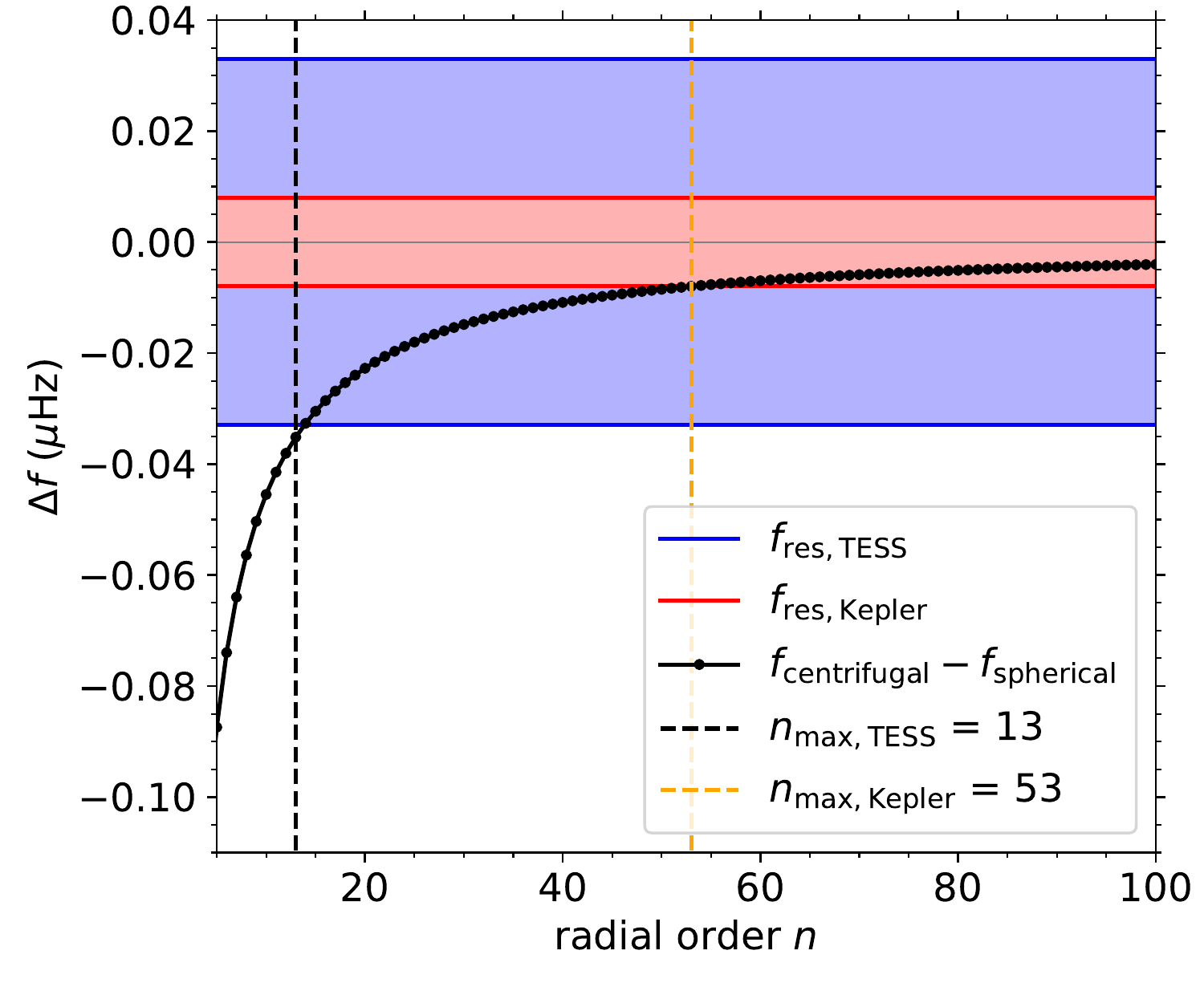}}
    \caption{Asymptotic frequency differences (black) for dipole $(k=0,m=1)$ modes at $\Omega =0.15\,\Omega_{\mathrm{c}}$ for the central $2.0\,\mathrm{M}_{\odot},$ $X_{\mathrm{c}}=0.50$ \texttt{MESA} model. Red and blue bands represent the frequency resolution of \textit{Kepler} and TESS respectively. The respective values of $n_{\mathrm{max}}$ are indicated by vertical dashed lines.}
    \label{fig:detect_central_dipole}
\end{figure}

To quantify the effect the centrifugal deformation of the star on the pulsation frequencies, we computed the frequency differences between asymptotic frequencies calculated in the classical formulation of the TAR, and those calculated in the generalised formulation (through Eq.\,\ref{eq:asymptotic_frequencies}). We consider these frequency differences as a function of the radial order $n$, as this is similar to the common diagnostic observables used in g-mode asteroseismic modelling \citep{aerts2018}. We compare the obtained frequency differences with the frequency resolutions ($1/T_{\mathrm{obs}}$) of \textit{Kepler} and TESS light curves covering  quasi-continuously observation times of $T_{\mathrm{obs}} = 4\,$years and $T_{\mathrm{obs}} = 351\,$days, respectively. In this way, we were able to deduce the radial orders $n_{\mathrm{max}}$ for which the frequency differences are expected to be detectable in the absence of instrumental effects and assuming excellent knowledge of the equilibrium models representing an observed star. In reality asteroseismic modelling never represents the stellar oscillations perfectly, meaning that the optimal reported $n_{\mathrm{max}}$ leading to the largest frequency differences due to the centrifugal deformation are in fact lower limits of detectability in real applications of asteroseismology. The results for prograde dipole $(k=0,m=1)$ modes in the central $(2.0\,\mathrm{M}_{\odot}, X_{\mathrm{c}}=0.50)$ model rotating at $0.15\,\Omega_{\mathrm{c}}$ are displayed in Fig.\,\ref{fig:detect_central_dipole}. These computations were done for all the models in our grid and the considered modes. The results are listed in Table\,\ref{table:detectability}.

\begin{figure}
    \centering
    \resizebox{\hsize}{!}{\includegraphics{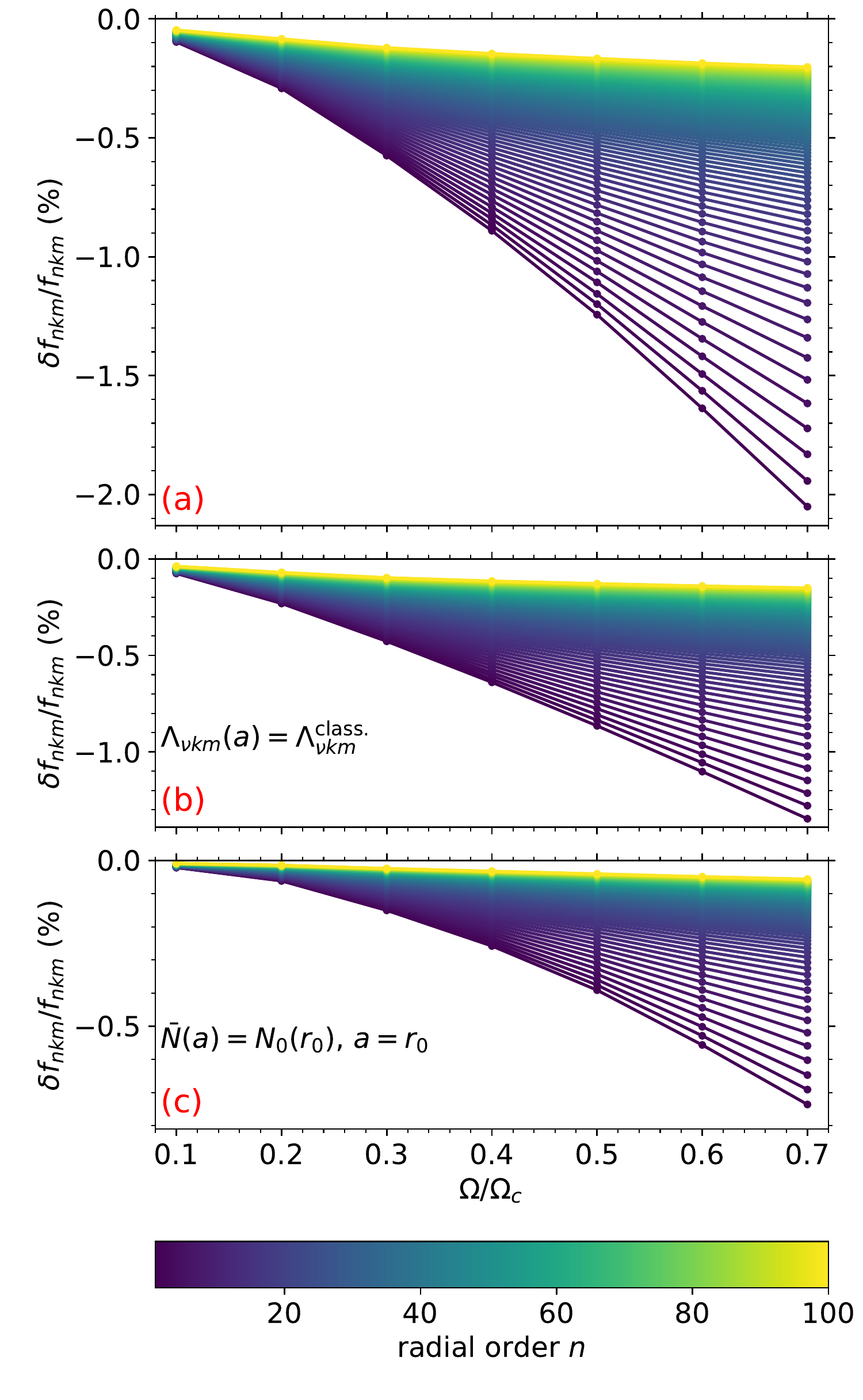}}
    \caption{\emph{(a)} Fractional frequency difference $\delta f_{nkm}/f_{nkm}$\,, with $\delta f_{nkm} = f_{\mathrm{centrifugal}} - f_{\mathrm{spherical}}$ and $f_{nkm}=f_{\mathrm{centrifugal}}$ in function of the fractional rotation rate, for $n=1$ to $n=100$\,. Calculations based on central $2\,\mathrm{M}_{\odot}$, $X_{\mathrm{c}}=0.50$ \texttt{MESA} model with $\Omega/\Omega_{\mathrm{c}} \in [0.10-0.70]$, exhibiting prograde dipole sectoral $(k=0,m=1)$ modes. \emph{(b)} Same as (a), but now with the isolated effect of the deformation of the Brunt-V\"ais\"al\"a frequency profile and radial coordinate (achieved by setting $\Lambda_{\nu km}(a) = \Lambda_{\nu km}^{\mathrm{class.}}$). \emph{(c)} Same as (a), but now with the isolated effect of the deviation of the eigenvalue of the GLTE (achieved by setting $\bar{N}(a)=N_{0}(r_0)$ and $a=r_0$).}
    \label{fig:applicability}
\end{figure}

\begin{table}
\caption{Potential detectability of the effect of the centrifugal acceleration in space-based photometric observations.}
\label{table:detectability}
\centering
\begin{tabular}{l c c c}
\hline\hline
model & $f_{\mathrm{rot}}$ & $n_{\mathrm{max}}$  & $n_{\mathrm{max}}$ \\
& [$\mu$Hz] & (\textit{Kepler}) & (TESS) \\
\hline
central model & & & \\ 
$2.0\,\mathrm{M}_{\odot}$, $X_{\mathrm{c}} = 0.5$, $Z=0.014$, & 4.35 & 53 & 13 \\
$D_{\mathrm{mix}} = 1\,\mathrm{cm}^2/\mathrm{s}$, $f_{\mathrm{ov}}=0.015$\,, & & & \\
$0.15\,\Omega_{\mathrm{c}}$, $(k,m) = (0,1)$  & & & \\
\hline
$1.5\,\mathrm{M}_{\odot}$ & 4.87 & 42 & 10 \\
$3.0\,\mathrm{M}_{\odot}$ & 3.72 & 62 & 16\\
$5.0\,\mathrm{M}_{\odot}$ & 3.11 & 69 & 18\\
$X_{\mathrm{c}} = 0.7$ & 6.01 & 74 & 19\\
$X_{\mathrm{c}} = 0.3$ & 2.99 & 39 & 9\\ 
$X_{\mathrm{c}} = 0.1$ & 1.88 & 28 & 6\\
$Z=0.010$ & 4.62 & 60 & 15\\
$Z=0.018$ & 4.14 & 50 & 13\\
$D_{\mathrm{mix}} = 0.1\,\mathrm{cm}^2/\mathrm{s}$ & 4.35 & 53 & 13\\
$D_{\mathrm{mix}} = 10\,\mathrm{cm}^2/\mathrm{s}$ & 4.35 & 53 & 13\\
$f_{\mathrm{ov}}=0.005$ & 4.51 & 55 & 14\\
$f_{\mathrm{ov}}=0.030$ & 4.13 & 50 & 13\\
$0.10\,\Omega_{\mathrm{c}}$ & 2.90 & 31 & 7 \\
$0.20\,\Omega_{\mathrm{c}}$ & 5.80 & 82 & 22\\
$0.30\,\Omega_{\mathrm{c}}$ & 8.70 & $> 100$ & 39 \\
$0.40\,\Omega_{\mathrm{c}}$ & 11.59 & $> 100$ & 59 \\
$0.50\,\Omega_{\mathrm{c}}$ & 14.49 & $> 100$ & 82 \\
$0.60\,\Omega_{\mathrm{c}}$ & 17.40 & $> 100$ & $> 100$ \\
$k=0,\,m=2$ & 4.35  & $> 100$ & 25 \\
\hline
$(k,m) = (-2,-1)$, $0.20\,\Omega_{\mathrm{c}}$ & 5.80 & \ldots & \ldots\\
$(k,m) = (-2,-1)$, $0.30\,\Omega_{\mathrm{c}}$ & 8.70 & 10 & \ldots \\
$(k,m) = (-2,-1)$, $0.40\,\Omega_{\mathrm{c}}$ & 11.59 & 14 & \ldots \\
$(k,m) = (-2,-1)$, $0.50\,\Omega_{\mathrm{c}}$ & 14.49 & 85 & 5 \\
$(k,m) = (-2,-1)$, $0.60\,\Omega_{\mathrm{c}}$ & 17.39 & $> 100$ & 8 \\
$(k,m) = (-2,-1)$, $0.70\,\Omega_{\mathrm{c}}$ & 20.29 & $> 100$ & 9 \\
\hline
\end{tabular}
\tablefoot{The first entry of the table (above the line) contains the result for the central model, of which the \texttt{MESA} input parameters, fractional rotation rate and mode identification are given. Subsequently, each of the indicated parameters/rotation rates/mode identifications is varied with respect to the central model. The influence of the rotation rate is evaluated again separately for $(k,m) = (-2,-1)$.}
\end{table}

To assess whether the centrifugal acceleration is a necessary ingredient in asteroseismic modelling, we computed the fractional differences between asymptotic frequencies in centrifugally deformed and spherically symmetric stars for a range of rotation rates $\Omega/\Omega_{\mathrm{c}} \in [0.10-0.70]$. For these calculations, the central $2\,\mathrm{M}_{\odot}$, $X_{\mathrm{c}}=0.5$ equilibrium model was used. We restricted these tests to prograde dipole sectoral $(k=0,m=1)$ modes, as these are most frequently observed. The results are displayed in Fig.\,\ref{fig:applicability}. The middle and bottom panel of this figure show that the deformation of the stellar structure and the divergence of the eigenvalues of the GLTE each cause an increase of $\delta f_{nkm}/f_{nkm}$ (in absolute value) for increasing  $\Omega=0.1\,\Omega_{\mathrm{c}}$ to $0.7\,\Omega_{\mathrm{c}}$. The fractional frequency differences caused by both effects are on the order of $1\,\%$. And while the indirect effect of the deformed stellar structure has a slightly larger impact, this again shows that both the deformation of the stellar structure itself and the generalisation of the TAR are required to accurately describe the effect of the centrifugal acceleration on GIW.

The frequency differences provide an assessment for the applicability of the analytical framework in \citet{mathis+prat2019} in terms of the fraction of the critical rotation rate $\Omega_{\mathrm{c}}$. For rotation rates near the critical value, i.e. $\Omega/\Omega_{\mathrm{c}} = 0.80-0.99$ (not shown in Fig.\,\ref{fig:applicability}) the behaviour of $\delta f_{nkm}/f_{nkm}$ deviates from the smooth curves in the figure for the lower values. This indicates that the assumptions in the analytical frame work (e.g., only taking $l=0$ and $l=2$ projections into account) no longer hold for higher rotation rates than those shown in Fig.\,\ref{fig:applicability}. The case where the centrifugal acceleration is treated in a non-perturbative manner will be part of forthcoming work (Dhouib et al., in prep.).

%%%%%%%%%%%%%%%%%%%%%%%%%%%%%%%%% DISCUSSION & CONCLUSIONS%%%%%%%%%%%%%%%%%%%%%%%%%%%%%%%%%
\section{Discussion \& conclusions}\label{sect:discussion}
The results in the previous section reveal that the  centrifugal deformation of the star implies a shift in the extrema of the eigenfunctions towards the equator of the star. This effect becomes more outspoken towards the surface, since the centrifugal acceleration (and hence the dimensionless deformation factor $\varepsilon$) becomes more important as one moves from the centre to the outer layers. The eigenfunctions centred around the equator reveal narrower maxima, while the extrema of the off-equator eigenfunctions move inwards, towards the equator.
We thus find that the effect of the centrifugal acceleration  modifies the one of the Coriolis acceleration included in the CLTE. More precisely, the Coriolis acceleration is responsible for a concentration of GIW towards the equator, while the centrifugal acceleration causes these equatorial bands to become narrower and the off-equator extrema of eigenfunctions shift inwards.

We found differences between the proof-of-concept study by \citet{mathis+prat2019} and the quantitative numerical results in this work. The current new implementation uses 
a different approach for the computational aspects to solve the GLTE. In order to avoid numerical issues, \citet{mathis+prat2019} made use of linearized expressions for the coefficients of the GLTE.
However, such an approach revealed the solutions to be 
rather dependent on the used number of collocation points. For this reason, we circumvented the use of numerical derivatives to solve the GLTE analytically where possible,
as described in Sect.\,\ref{sect:solver_for_GLTE}.
In this way, the avoided crossings in the solution spectrum for the GLTE (Fig.\,\ref{fig:spectrum}; their Figs.\,5\,\&\,6) are circumvented here. Moreover, the current numerical approach leads to relatively modest shift of the eigenfunctions towards the equator 
as shown graphically in our Figs.\,\ref{fig:k0m2nu2Om30_hough},\,\ref{fig:other_houghs}\,\&\,\ref{fig:hough_rotation}, as an improvement compared to the more pronounced behaviour found in  Figs.\,7\,\&\,8 in  \citet{mathis+prat2019}.

In essence, the centrifugal acceleration affects two aspects within the theoretical description of g-mode pulsators. On the one hand, the stellar structure as a whole becomes deformed, which naturally leads to changes in the Brunt-V\"ais\"al\"a frequency profile and therefore in the cavities of the gravity modes. This is an indirect effect. On the other hand, the g-mode oscillation equations, in the form of the Laplace tidal equations, are altered and get a radial dependence, propagating into deformed solutions compared to those of the CLTE.  This is a direct effect. In the case of prograde sectoral modes, these aspects have similar effects on the mode frequencies, as demonstrated in Sect.\,\ref{sect:asymptotic_period_spacing_patterns}.

The detectability of the effect of the centrifugal acceleration in space-based photometry decreases with increasing stellar age (decreasing $X_{\mathrm{c}}$), increasing metallicity $Z$ and increasing core overshooting $f_{\rm ov}$. Further, as seen in in Table\,\ref{table:detectability}, the value of $n_{\mathrm{max}}$ increases with increasing stellar mass and increasing rotation rate, since the deformation factor and rotation rate scale as $\varepsilon \sim \Omega^2$.

Comparing the $n_{\mathrm{max}}$ values with radial-order distributions of observed GIWs \citep[and Pedersen et al., submitted]{li2020}, we conclude that  the values we obtained here correspond well with  the observations. This implies that it should be possible to detect differences between theoretically computed pulsation frequencies assuming spherically symmetric versus deformed stellar models as computed in this work for real stars, provided that the comparison is done for models with the same input physics and rotation rate.
The frequency differences as shown in Fig.\,\ref{fig:detect_central_dipole} are of similar order as typical uncertainties of the observed frequencies used in forward asteroseismic modelling \citep{aerts2018}.

Frequency differences caused by other approximations that are commonly made in the TAR, such as the Cowling approximation and the neglect of the horizontal component of the rotation vector, are smaller than or comparable to those caused by the centrifugal acceleration. This is shown in Appendix\,\ref{app:approx-eval}, where we obtain relative differences $\lesssim 1\,\%$ dependent on the stellar rotation rate, comparable to the differences between the TAR and the \citet{mathis+prat2019} framework. A comparison between the TAR and full 2D-computations carried out with \texttt{ACOR} by \citet[Figure 1]{ouazzani2017}, gave similar results. This indicates that, just like the neglect of the centrifugal acceleration, other assumptions made within the TAR also have negligible or minor effects on the pulsation calculations. A limitation of the asymptotic expressions that is still present in the \citet{mathis+prat2019} framework, is that the effect of the centrifugal acceleration increases with decreasing radial order of the $g$ modes. At low radial order, both the TAR and the asymptotic expressions cease to be valid.

In practice,  Fig.\,\ref{fig:applicability} reveals that the fractional frequency differences due to the centrifugal acceleration remain well below $1\,\%$ for high-order g~modes in the asymptotic regime. 
Comparing this with typical fractional frequency differences in \citet[][Table 2]{aerts2018}, 
points out that the effect of the centrifugal acceleration is negligible compared to that introduced by some aspects of missing input physics, such as 
atomic diffusion in slow rotators among the g-mode pulsators \citep{mombarg2020} or near-core boundary mixing in fast rotating single SPB stars \citep[and Pedersen et al., submitted]{moravveji2016,szewczuk2018} and SPB binaries \citep{johnston2019}. Hence, although the frequency differences induced by the centrifugal acceleration are expected to be detectable, one can ignore them for initial attempts of asteroseismic modelling. Once appropriate equilibrium models are found that explain well the overall structure in the measured period spacing pattern of a star, it is meaningful to test the effect of the centrifugal acceleration for those models to see if it brings an improved fit, particularly for stars with relatively fast rotation.

Following the conclusions of \citet{mathis+prat2019} and ours, a next logical step would be to include the effect of the centrifugal acceleration in stellar pulsation codes such as \texttt{GYRE}. Although our work has shown that the centrifugal acceleration can be treated as a lower-priority ingredient for forward asteroseismic modelling compared to other missing ingredients in the input physics, its effect is in principle detectable at the level of the mode computations. Hence, including it would lead to overall more realistic stellar pulsation predictions. 

\begin{acknowledgements}
We thank the referee for their encouragement. We thank the \texttt{MESA} and \texttt{GYRE} developers for their efforts and for making their codes publicly available. The research leading to these results has received funding from the European Research Council (ERC) under the European Union’s Horizon 2020 research and innovation programme (grant agreements N$^\circ$670519: MAMSIE with PI Aerts and N$^\circ$647383: SPIRE with PI Mathis) and from the KU\,Leuven Research Council (grant C16/18/005: PARADISE). TVR is funded by the Research Foundation Flanders (FWO) by means of a Junior Postdoctoral Fellowship  under grant agreement N$^\circ$12ZB620N. VP and SM acknowledge support from the CNES PLATO grant at CEA/DAP.
\end{acknowledgements}

\bibliographystyle{aa}
\bibliography{centrifugaldeformation_arxiv}

\appendix
\section{Centrifugal deformation of Brunt-V\"ais\"al\"a frequency profiles}\label{app:BV_frequency}
After calculating the deformed stellar radius $r(a,\theta)$ using Eqs.\,(\ref{r_ifo_a}) and (\ref{eq:modal_amplitude_epsilon}), we compute the two-dimensional centrifugally deformed Brunt-V\"ais\"al\"a frequency profile
\begin{equation}
N^2(a,\theta) = -\frac{\bar{g}}{r}\left[\frac{\mathrm{d}\ln\bar{\rho}}{\mathrm{d}\ln r} - \frac{1}{\bar{\Gamma}_{1}}\frac{\mathrm{d}\ln \bar{P}}{\mathrm{d}\ln r}\right]\,,\label{eq:barN2_calc}
\end{equation}
to ensure numerical stability and to account for the nonlinear dependence of $N^2(a,\theta)$ on the density $\bar{\rho}$ and the pressure $\bar{P}$. These calculations are done at discrete values of the co-latitude $\theta$, and we use the calculation methods for $\bar{\rho}$ and $\bar{P}$ described in Section \ref{sect:deformation_stellar_structure}.

The perturbed adiabatic exponent $\bar{\Gamma}_{1}$ can then be estimated as follows:
\begin{align}
    \frac{1}{\bar{\Gamma}_{1}} &= \left(\frac{\partial \ln\bar{\rho}}{\partial \ln\bar{P}}\right)_{\bar{S}} = \left(\frac{\frac{\partial \ln\bar{\rho}}{\partial a}}{\frac{\partial \ln \bar{P}}{\partial a}}\right)_{\bar{S}} \\
    &\approx \left(\frac{\partial \ln\bar{\rho}}{\partial a}\, \frac{\mathrm{d} \ln P_{0}}{\mathrm{d} r_0}\right) \left(\frac{\mathrm{d} \ln \rho_{0}}{\mathrm{d} r_0}\,\frac{\partial \ln \bar{P}}{\partial a}\right)^{-1} \left(\frac{\partial \ln \rho_{0}}{\partial \ln P_{0}}\right)_{S} \label{gamma_assumption} \\
    &= \left(\rho_{0}\bar{P}\frac{\partial\bar{\rho}}{\partial a}\,\frac{\mathrm{d} P_{0}}{\mathrm{d} r_0}\right) \left(\bar{\rho} P_{0} \frac{\mathrm{d} \rho_{0}}{\mathrm{d} r_0}\,\frac{\partial \bar{P}}{\partial a}\right)^{-1} \frac{1}{\Gamma_{1,0}}\,,
\end{align}
where quantities $X$ from the spherical non-rotating input model are indicated as $X_0$.

The perturbed gravitational acceleration is the gradient of the \emph{effective} gravitational potential $\bar{\phi}$;
\begin{equation}
    \bar{g} = \vec{\bar{\nabla}}\bar{\phi} = \vec{\bar{\nabla}}\left(\phi_{0} + \phi_{1} + U\right) = g_{0} + \vec{\bar{\nabla}}(\phi_{1} + U)\,.
\end{equation}
The spheroidal gradient $\vec{\bar{\nabla}}$, defined as \citep{mathis+prat2019}
\begin{equation}\label{gradient}
    \vec{\bar{\nabla}}X \equiv\frac{\partial X}{\partial a} \vec{\widetilde{e}}_{a} + \frac{1}{a}\frac{\partial X}{\partial \theta}\vec{\widetilde{e}}_{\theta} + \frac{1}{a\sin\theta}\frac{\partial X}{\partial \varphi}\vec{\widetilde{e}}_{\varphi}\,,
\end{equation}
reduces to the total derivative $\mathrm{d}/\mathrm{d}a$ since the calculations of the perturbed quantities are discretized in the co-latitude $\theta$, i.e. they are calculated for a specific, constant value of $\theta$.

To avoid direct numerical differentiation of the perturbed pressure $\bar{P}$, it is rewritten as: 
\begin{align}
    \frac{\mathrm{d}\bar{P}}{\mathrm{d}r} &= -\bar{\rho}\frac{\mathrm{d}\bar{\phi}}{\mathrm{d}r} = -\frac{\bar{\rho}}{1+4\varepsilon}\frac{\mathrm{d}}{\mathrm{d}a}\bar{\phi} \\
    &= -\frac{\bar{\rho}}{1+4\varepsilon}\left[g_{0} + \frac{\mathrm{d}}{\mathrm{d}a}(\phi_{1} + U)\right] \\
    &= -\frac{\bar{\rho}}{1+4\varepsilon}\left[g_{0} + \frac{\mathrm{d}}{\mathrm{d}r_0}(\phi_{1} + U)\right]\,,\label{eq:dPbardr}
\end{align}
where we map the radius $r_0$ of the non-deformed stellar model onto the pseudo-radius $a$, as in the evaluation of Eq.\,(\ref{eq:map_radco}). Within their respective coordinate systems, both $r_0$ and $a$ coincide with the isobaric radial coordinate. Consequently, only the last derivative with respect to $r_0$ in Eq.(\ref{eq:dPbardr}) remains to be evaluated through numerical differentiation. This is numerically more stable than performing the derivation of $\bar{P}$  with respect to $r$. Similarly, for the derivative of the perturbed density $\bar{\rho}$ we have:
\begin{align}
    \frac{\mathrm{d} \bar{\rho}}{\mathrm{d} r} &= \frac{1}{1+4\varepsilon}\frac{\mathrm{d}}{\mathrm{d} a}\left(\rho_{0} + \rho_{1}\right) \\
    &= \frac{1}{1+4\varepsilon}\frac{\mathrm{d}}{\mathrm{d} r_0}\left(\rho_{0} + \rho_{1}\right) \\
    &= \frac{1}{1 + 4\varepsilon}\left[\frac{\mathrm{d} \rho_{0}}{\mathrm{d} r_0} + \frac{\mathrm{d}}{\mathrm{d} r_0}\left(\frac{1}{g_{0}}\frac{\mathrm{d} \rho_{0}}{\mathrm{d} r_0}\left(\phi_{1} + U\right)\right)\right] \\
    \begin{split}
    &= \frac{1}{1+4\varepsilon}\frac{\mathrm{d} \rho_{0}}{\mathrm{d} r_0}\left[1 - \left(\frac{\phi_{1} + U}{g_{0}^2}\right)\frac{\mathrm{d} g_{0}}{\mathrm{d} r_0} + \frac{1}{g_{0}}\frac{\mathrm{d} }{\mathrm{d} r_0}(\phi_{1} + U)\right] \\&\quad + \frac{\phi_{1} + U}{g_{0}(1+4\varepsilon)}\frac{\mathrm{d}}{\mathrm{d} r_0}\left(\frac{\mathrm{d} \rho_{0}}{\mathrm{d} r_0}\right)\,.\end{split}
\end{align}
The last three derivatives with respect to $a$ are again evaluated through numerical differentiation without numerical issues.

Finally, $\mathrm{d}P_{0}/\mathrm{d}r_0$ and $\mathrm{d}\rho_{0}/\mathrm{d}r_0$ can be expressed in terms of physical quantities included in the (non-deformed) \texttt{MESA} stellar structure profiles. The former can be calculated by using hydrostatic equilibrium (in a spherically symmetric star):
\begin{equation}
    \frac{\mathrm{d}P_{0}}{\mathrm{d}r_0} = -\rho_{0}g_{0},
\end{equation}
and the latter can be retrieved from the non-perturbed squared Brunt-V\"ais\"al\"a frequency:
\begin{equation}
    \frac{\mathrm{d}\rho_{0}}{\mathrm{d}r_0} = \rho_{0}\left(\frac{1}{\Gamma_{1,0}P_{0}}\frac{\mathrm{d}P_{0}}{\mathrm{d}r_0} - \frac{N_{0}^2}{g_{0}}\right)\,.
\end{equation}

\section{Computation of asymptotic frequencies}\label{app:asymptotic}
\begin{figure}
    \centering
    \resizebox{0.8\hsize}{!}{\includegraphics{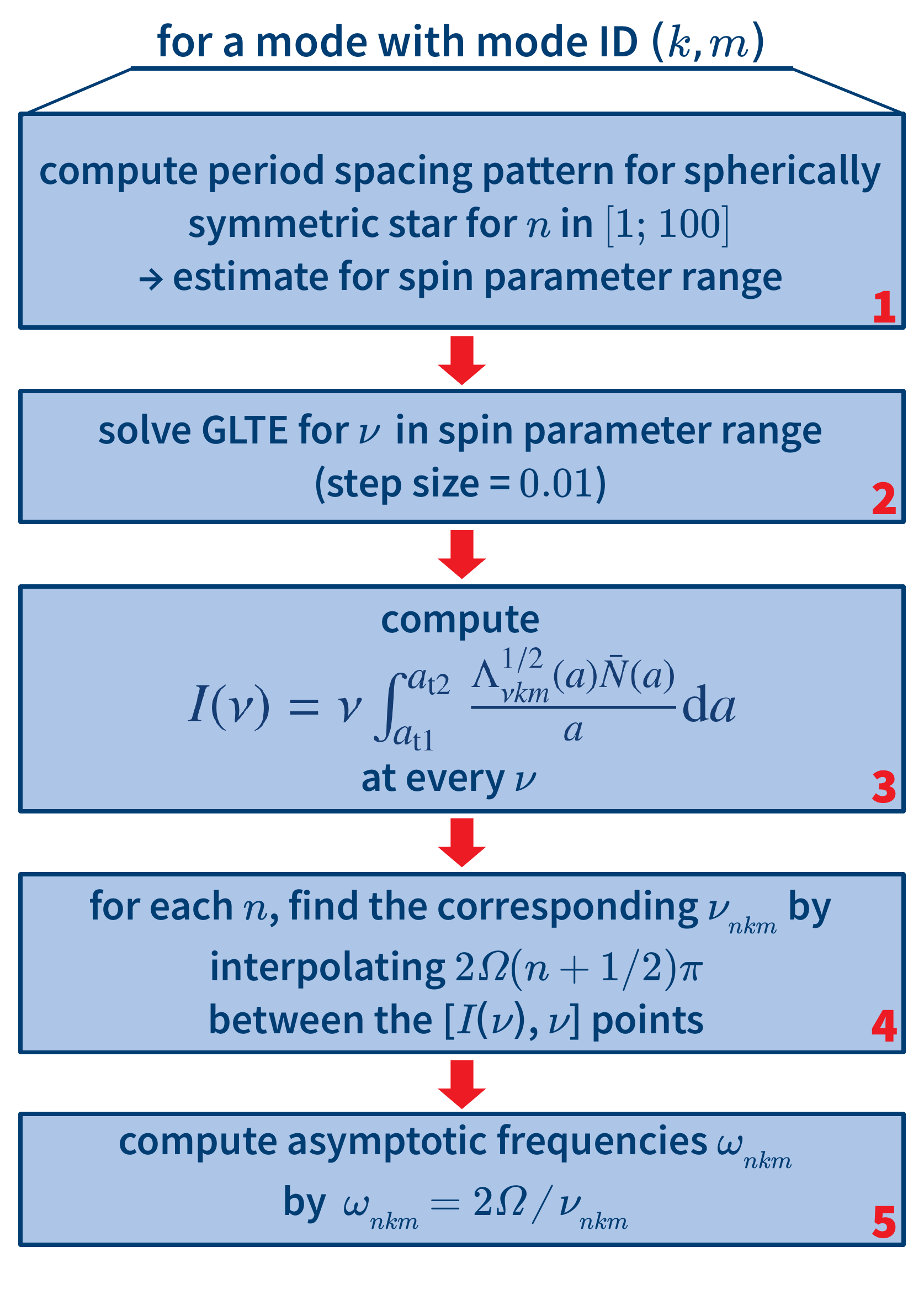}}
    \caption{Flowchart describing the strategy used in this work to compute asymptotic pulsation frequencies in a centrifugally deformed star.}
    \label{fig:flowchart}
\end{figure}

Here we provide our strategy to compute the asymptotic frequencies and period spacing patterns presented in Sect.\,\ref{sect:asymptotic_period_spacing_patterns}\,\&\,\ref{sect:detectability}. This is summarised in a flowchart in Fig.\,\ref{fig:flowchart} and concerns the computation of the asymptotic frequencies for given \emph{pulsation mode} with mode identification $(k,m)$, for a chosen \emph{stellar structure model} with a particular rotation rate (in terms of $\Omega_{\mathrm{c}}$).

The period spacing pattern for a spherically symmetric, uniformly rotating star is computed within a radial order range of $n=1$ to $n=100$. This particular range is motivated by the radial order distributions of typically observed GIWs \citep{li2020}. Based on the resulting period spacing pattern, a suitable $\nu-$range is chosen. To anticipate the (relatively small) effect of the centrifugal force on the asymptotic pulsation periods, small margins are taken above and below this range (typically $\pm 0.5$).
In the next step, the GLTE is solved for a number of spin parameters within the predetermined range. A step size of 0.01 in $\nu$ is found suitable to avoid numerical inaccuracies, while maintaining reasonable computation times (typical values for $\nu_{\mathrm{max}}-\nu_{\mathrm{min}}$ are on the order of $10$).
In the subsequent step, the expression for the asymptotic pulsation frequencies Eq.\,(\ref{eq:asymptotic_frequencies}) is rewritten as follows:
\begin{equation}
\frac{1}{\omega_{nkm}} = \frac{(n+1/2)\pi}{\displaystyle{\int_{a_{1}}^{a_{2}} \frac{\Lambda_{\nu km}^{1/2}(a)\bar{N}(a)}{a}\mathrm{d} a}}\,,\label{eq:cofreq-pattern}
\end{equation}
and multiplied by $2\,\Omega$ to arrive at:
\begin{equation}\label{eq:interp_eq}
    I(\nu) = \nu\int_{a_{1}}^{a_{2}} \frac{\Lambda_{\nu km}^{1/2}(a)\bar{N}(a)}{a}\mathrm{d} a = 2\Omega(n+1/2)\pi\,.
\end{equation}
The LHS of this equation is then evaluated at every $\nu$ in the spin parameter range, while the RHS is evaluated for every $n\in [1;100]$.
In the fourth step, the spin parameters $\nu_{nkm}$ corresponding to the radial orders $n\in [1;100]$ are computed through linear interpolation of the RHS of Eq.\,(\ref{eq:interp_eq}) to the $\left(I(\nu), \nu\right)$-points from the previous step.
Finally, the asymptotic angular frequencies $\omega_{nkm}$ are retrieved by taking the inverse of the $\nu_{nkm}$ from the previous step, and multiplying them with $2\Omega$. The corresponding asymptotic pulsation periods are $P_{nkm}=2\pi/\omega_{nkm}$. The corresponding pulsation periods in the inertial (observer's) frame $P_{nkm}^{\mathrm{in}}$ are then found through \citep{bouabid2013}:
\begin{equation}
    P_{nkm}^{\mathrm{in}} = \frac{P_{nkm}}{1+m\frac{P_{nkm}}{P_{\mathrm{rot}}}}\,,
\end{equation}
with $P_{\mathrm{rot}}$ the rotation period.

\section{Comparison with other theoretical assumptions}
\label{app:approx-eval}
Other assumptions that are included within the TAR, are still made in the generalised framework developed by \citet{mathis+prat2019}, such as the neglect of the horizontal rotation vector component $\vec\Omega_{\rm H}$ and the Cowling approximation. Here we assess the impact of these approximations on our results.

\subsection{Neglect of the horizontal rotation vector} 
The spheroidal radial component of the momentum equation is given by \citet[Eq.(13)]{mathis+prat2019} as
\begin{align}
\begin{split}
 &-\bar{N}^2\left(\frac{\omega}{\bar{N}}\right)^2\left[\left(1 + 2\left(\varepsilon + a\partial_a\varepsilon\right)\right)\xi_a + \xi_\theta\partial_\theta\varepsilon\right] \\
 &\qquad- i \bar{N}^2\left(\frac{\omega}{\bar{N}}\right)\left(\frac{2\Omega}{\bar{N}}\right)\left(1 + 2\varepsilon + a\partial_a\varepsilon\right)\sin\theta\,\xi_\varphi\\
 &\qquad= -\partial_a\widetilde{W} - \bar{N}^2\xi_a - \frac{1}{\bar{\rho}^2}\partial_a\bar{\rho}\widetilde{P}   \label{eq:mom-eq}
\end{split}\,,
\end{align}
where $\widetilde{W} = \widetilde{P}/\bar{\rho}$ with $\widetilde{P}$ the wave fluctuation of the pressure, and $\vec\xi = (\xi_a, \xi_\theta, \xi_\phi)$ is the Lagrangian displacement vector. Within the generalised TAR-framework, the two terms on the left-hand side (LHS) of Eq.(\ref{eq:mom-eq}) are neglected in favour of the $\bar{N}^2\xi_a$-term on the right-hand side (RHS), and the second LHS-term contains the horizontal component of the rotation vector $\Omega_{\rm H} = \Omega\sin\theta$. \citet{mathis+prat2019} combined the simplified Eq.(\ref{eq:mom-eq}) with the horizontal components of the momentum equation, and solved the resulting system for a selected pulsation mode identification $(k,m)$ and spin parameter value $\nu$ as a function of the normalised pressure $W^\prime_{\nu km}(a,\theta)$:
\begin{align}
    \xi^\prime_{a;\nu km}(a,\theta) &= -i\,k_{V;\nu km}(a)\frac{W^\prime_{\nu km}}{\bar{N}^2 (a)} = -i\frac{\sqrt{\Lambda_{\nu km}(a)}}{a\,\omega_{km}}\frac{W^\prime_{\nu km}}{\bar{N}(a)}\label{eq:xi_a}\\
    \xi^\prime_{\theta;\nu km}(a,\theta) &= \frac{1}{a}\frac{1}{\omega_{km}^2}\frac{1}{\mathcal{D}}\left[\partial_\theta W^\prime_{\nu km} - m\nu\frac{\cos\theta}{\sin\theta}C W^\prime_{\nu km}\right]\label{eq:xi_th}\\
    \xi^\prime_{\varphi;\nu km}(a,\theta) &= i\frac{1}{a}\frac{1}{\omega_{km}^2}\frac{1}{\mathcal{D}}\left[\nu C \cos\theta \partial_\theta W^\prime_{\nu km} - \frac{m}{\sin\theta}W^\prime_{\nu km}\right]\label{eq:xi_phi}
\end{align}
Expressions for the coefficients $C\left(a,\theta\right)$ and $D\left(a,\theta\right)$ are given in Eqs.(\ref{eq:dxvareps}) to (\ref{eq:evar}).

\begin{figure}
    \centering
    \includegraphics[width=88mm]{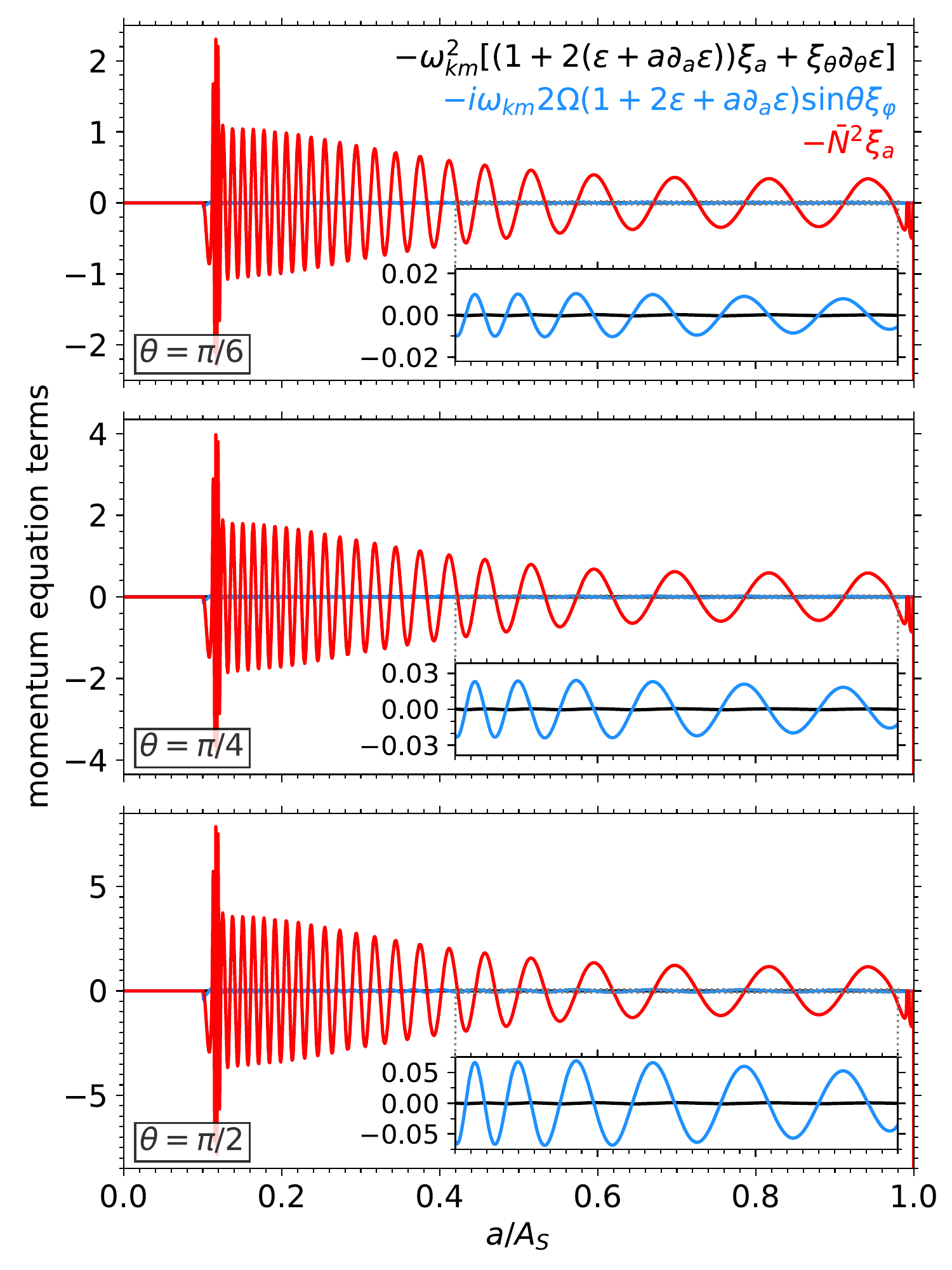}
    \caption{The $\bar{N}^2\xi_a$-term (red), the first LHS-term (black) and second LHS-term (blue) in the momentum equation Eq.(\ref{eq:mom-eq}), calculated for the $(n,k,m) = (50,0,1)$ pulsation mode in the central model of our \texttt{MESA} grid, with $2\,\mathrm{M}_{\odot}$, $X_{\mathrm{c}}=0.50$ and a rotation rate $\Omega = 0.15\,\Omega_{\mathrm{c}}$. The values are shown at co-latitudes $\theta$ = $\pi/6$ (top), $\pi/4$ (middle) and $\pi/2$ (bottom). Parts of the LHS-terms are shown again in the insets, magnified $27\times$.}
    \label{fig:momeq-terms}
\end{figure}

\begin{figure}
    \centering
    \includegraphics[width=88mm]{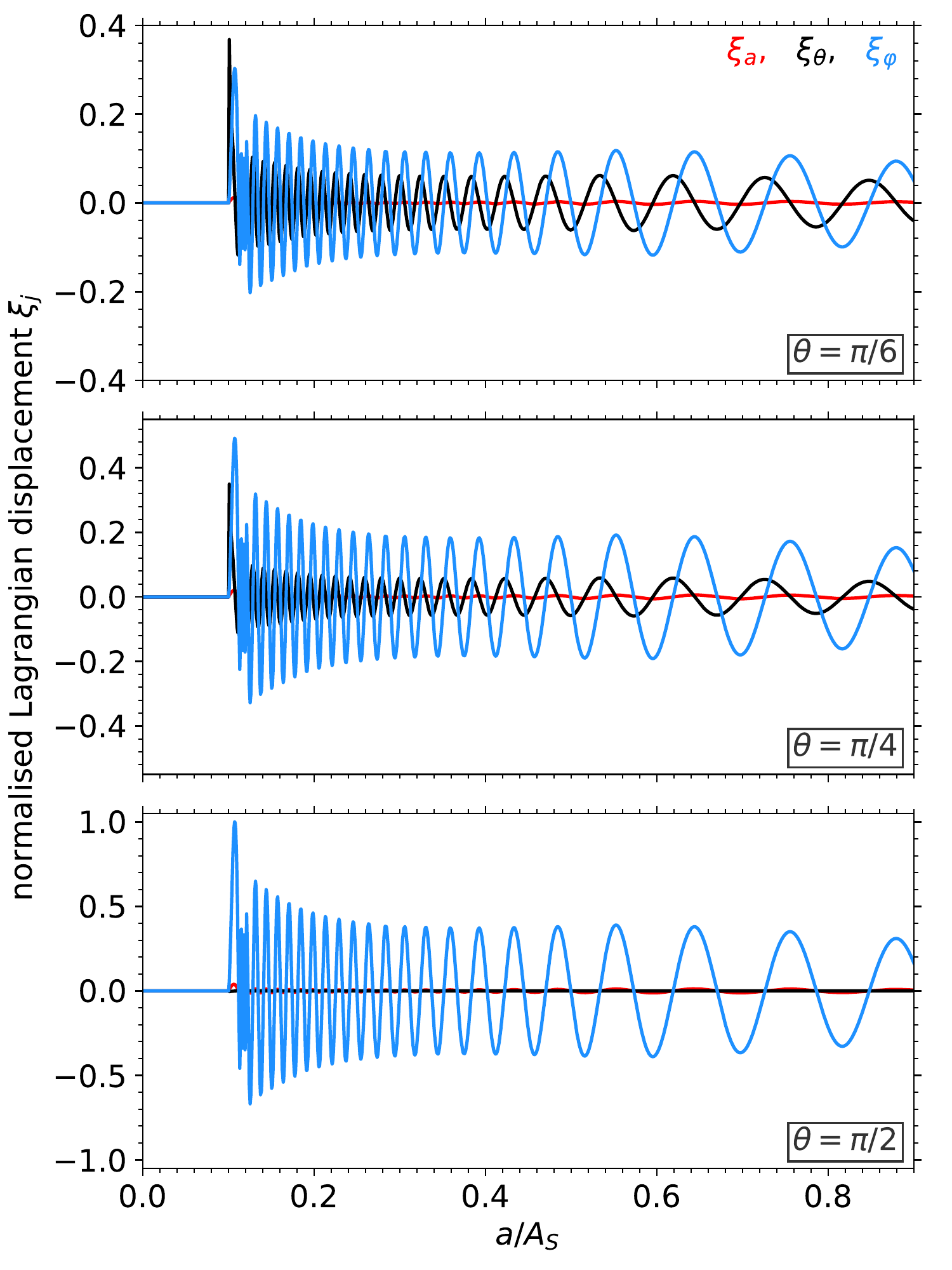}
    \caption{The (normalised) components of the Lagrangian displacement $\xi_a$ (red), $\xi_\theta$ (black) and $\xi_\phi$ (blue), calculated in the numerical simulations shown in Fig.\,\ref{fig:momeq-terms}. The functions are shown at co-latitudes $\theta$ = $\pi/6$ (top), $\pi/4$ (middle) and $\pi/2$ (bottom).}
    \label{fig:eigenfun}
\end{figure}

We assess the validity of neglecting the LHS-terms in Eq.(\ref{eq:mom-eq}) by taking their ratios with the $\bar{N}^2\xi_a$-term and filling in the solution for $\xi^\prime_{\nu km}$. For the first LHS-term, we obtain
\begin{align}
\begin{split}
 &\bar{N}^2\left(\frac{\omega_{km}}{\bar{N}}\right)^2\left[\left(1 + 2\left(\varepsilon + a\partial_a\varepsilon\right)\right)\xi_{a;\nu km} + \xi_{\theta;\nu km}\partial_\theta\varepsilon\right] \left[\bar{N}^2\xi_{a;\nu km}\right]^{-1}\\
 &\simeq \left(\frac{\omega_{km}}{\bar{N}}\right)^2\left(1 + 8\varepsilon\right) + \left(\frac{\omega_{km}}{\bar{N}}\right)\frac{3\varepsilon_{l=2}}{\mathcal{D}\sqrt{\Lambda_{\nu km}}}\\
 &\qquad\qquad\qquad\quad \times\left[\frac{\sin2\theta\,\partial_\theta W^\prime_{\nu km}}{2iW^\prime_{\nu km}} - \frac{m\nu C \cos^2\theta W^\prime_{\nu km}}{iW^\prime_{\nu km}}\right]\label{eq:LHS1}
\end{split},
\end{align}
and for the second LHS-term, we find
\begin{align}
\begin{split}
 &i \bar{N}^2\left(\frac{\omega_{km}}{\bar{N}}\right)\left(\frac{2\Omega}{\bar{N}}\right)\left(1 + 2\varepsilon + a\partial_a\varepsilon\right)\sin\theta\,\xi_\varphi \left[           \bar{N}^2\xi_{a;\nu km}\right]^{-1}\\
 &\simeq 
    \left(\frac{2\Omega}{\bar{N}}\right)\frac{\left(1 + 5\varepsilon\right)}{2\mathcal{D}\sqrt{\Lambda_{\nu km}}}\left[sin\left(2\theta\right)\nu \mathcal{C}\frac{\partial_\theta W^\prime_{\nu km}}{iW^\prime_{\nu km}} - \frac{2m W^\prime_{\nu km}}{iW^\prime_{\nu km}}\right]\label{eq:LHS2}
\end{split}.
\end{align}
Hence, the errors introduced by neglecting the terms in Eq.\ref{eq:mom-eq}, scale with $\omega_{km}/\bar{N}$ and $2\Omega/\bar{N}$, respectively. We can ignore the contribution from the horizontal component of the rotation vector when the GIWs propagate in a strongly stratified radiative region, i.e., for $2\Omega \ll \bar{N}$. This is further illustrated in Fig.\,\ref{fig:momeq-terms}, where we compare the terms in the momentum equation Eq.(\ref{eq:mom-eq}) for a $(n,k,m) = (50,0,1)$ pulsation mode in the central model of our MESA grid, with a rotation rate $\Omega = 0.15\,\Omega_{\mathrm{c}}$. The relative contribution of the neglected LHS-terms, which include the $\Omega\sin\theta$-component of the rotation vector, is $\sim\,1\,\%$. The (normalised) components of the Lagrangian displacement $(\xi_a, \xi_\theta, \xi_\phi)$, calculated in this simulation using Eqs.\,(\ref{eq:xi_a}) to (\ref{eq:xi_phi}) are shown in Fig.\,\ref{fig:eigenfun}. As can be seen, $\xi_a \ll \xi_\theta,\xi_\varphi$.

\subsection{The Cowling approximation}
\begin{figure}
    \centering
    \includegraphics[width=88mm]{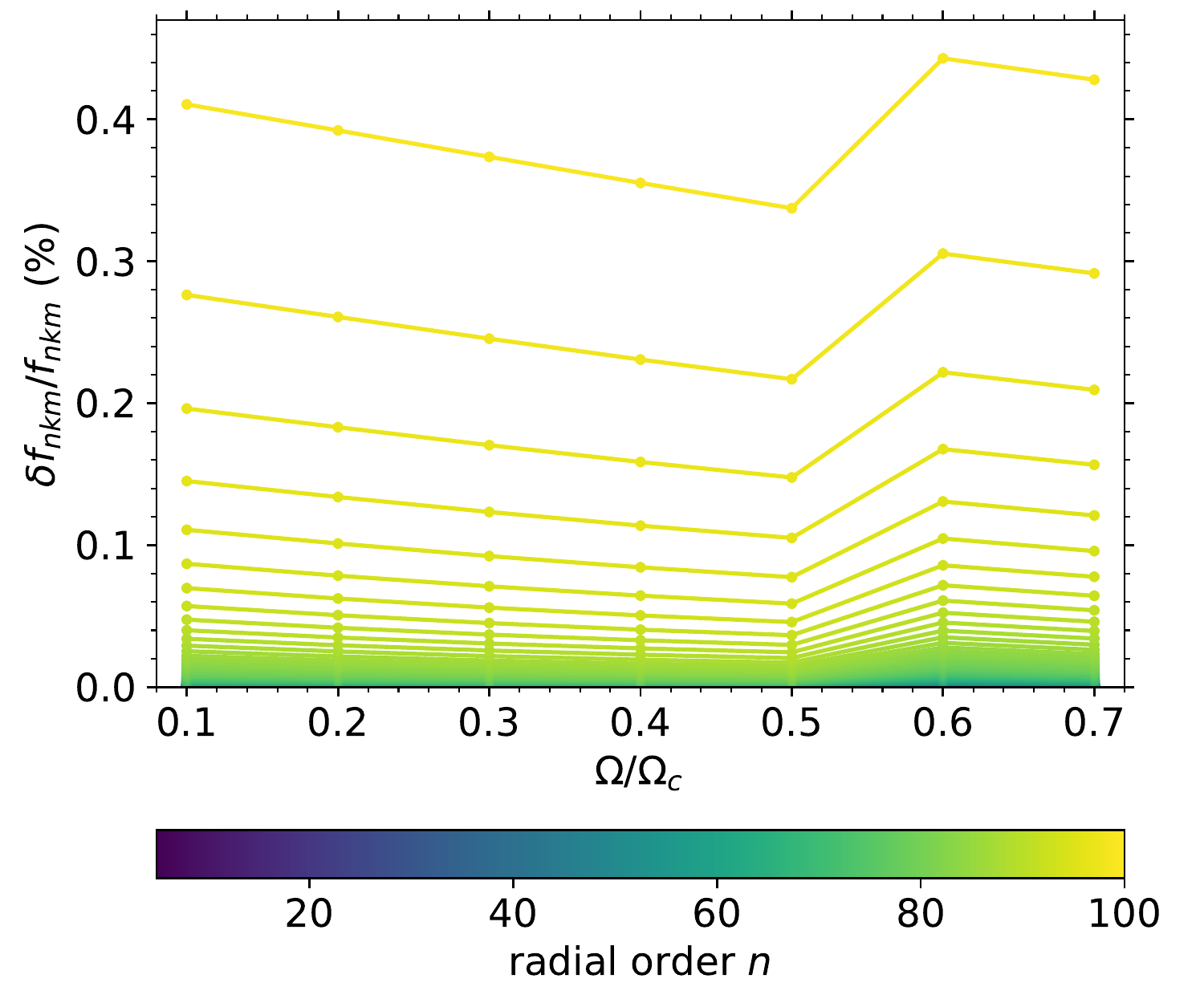}
    \caption{Relative pulsation frequency shifts $\delta f_{nkm}/f_{nkm}$ caused by the Cowling approximation, with $\delta f_{nkm} = f_{\rm Cowling} - f_{\rm without Cowling}$ and $f_{nkm} = f_{\rm without Cowling}$, as a function of the fractional rotation rate, for $n=1$ to $n=100$. Calculations are done using the central $2\,M_\odot$, $X_c=0.50$ \texttt{MESA} model with $\Omega/\Omega_{\mathrm{c}} \in \left[0.10 - 0.70\right]$, for prograde dipole sectoral $(k=0,m=1)$ modes.}
    \label{fig:cowling}
\end{figure}

We evaluate the Cowling approximation for the central \texttt{MESA} model in our grid ($2\,\mathrm{M}_{\odot}$, $X_{\mathrm{c}}=0.50$) using \texttt{GYRE}, for rotation rates $\Omega/\Omega_{\mathrm{c}} \in \left[0.10 - 0.70\right]$, without the centrifugal acceleration. As shown in Fig.\,\ref{fig:cowling}, the relative differences between the pulsation frequencies, caused by the Cowling approximation, are $\propto\,0.1\,\%$. These are smaller than the relative differences introduced by the centrifugal acceleration, but increase with increasing radial order of the \emph{g}~modes.

\end{document}